\pdfoutput=1
\documentclass[compsoc,conference,a4paper,10pt,times]{IEEEtran}
\IEEEoverridecommandlockouts
\usepackage{cite}
\usepackage{amsmath,amssymb,amsfonts}
\usepackage{algorithmic}
\usepackage{graphicx}
\usepackage{textcomp}
\usepackage{bmpsize}
\usepackage{xcolor}
\usepackage{lipsum}
\usepackage{graphicx}
\usepackage{makecell}
\usepackage[all]{nowidow}
\usepackage{enumitem}

\usepackage[colorlinks=true,urlcolor=black]{hyperref}

\PassOptionsToPackage{normalem}{ulem}
\usepackage{ulem}

\providecolor{added}{rgb}{0,0,1}
\providecolor{deleted}{rgb}{1,0,0}

\newcommand{\added}[1]{{\color{black}{}#1}}
\newcommand{\deleted}[1]{{}}

\newcommand{\lastadded}[1]{{\color{black}{}#1}}
\newcommand{\lastdeleted}[1]{{}}

\usepackage{xurl}

\usepackage[flushleft]{threeparttable}
\usepackage{comment}

\def\BibTeX{{\rm B\kern-.05em{\sc i\kern-.025em b}\kern-.08em
    T\kern-.1667em\lower.7ex\hbox{E}\kern-.125emX}}

\newcommand{\abbas}[1]{\textcolor{black}{#1}}
\newcommand{\ege}[1]{\textcolor{black}{#1}}
\newcommand{\egee}[1]{\textcolor{black}{#1}}

\usepackage{color, colortbl}
\usepackage{listings}
\usepackage{color}
\let\origthelstnumber\thelstnumber
\makeatletter
\newcommand*\Suppressnumber{%
  \lst@AddToHook{OnNewLine}{%
    \let\thelstnumber\relax%
     \advance\c@lstnumber-\@ne\relax%
    }%
}

\newcommand*\Reactivatenumber{%
  \lst@AddToHook{OnNewLine}{%
   \let\thelstnumber\origthelstnumber%
   \advance\c@lstnumber\@ne\relax}%
}
\definecolor{light-gray}{gray}{0.90}
    
\begin{document}

\title{SoK: Cryptojacking Malware}


\author{\IEEEauthorblockN{Ege Tekiner\IEEEauthorrefmark{1}, Abbas Acar\IEEEauthorrefmark{1}, A. Selcuk Uluagac\IEEEauthorrefmark{1}, Engin Kirda\IEEEauthorrefmark{2}, and Ali Aydin Selcuk\IEEEauthorrefmark{3}}
\IEEEauthorblockA{ \IEEEauthorrefmark{1}Florida International University, Email:\{etekiner, aacar001, suluagac\}@fiu.edu  \\  \IEEEauthorrefmark{2}Northeastern University, Email: \{ek\}@ccs.neu.edu  \\  \IEEEauthorrefmark{3}TOBB University of Economics and Technology, Email: \{aselcuk\}@etu.edu.tr }}

\maketitle

\begin{abstract}
Emerging blockchain and cryptocurrency-based technologies are redefining the way we conduct business in cyberspace. Today, a myriad of blockchain and cryptocurrency systems, applications, and technologies are widely available to companies, end-users, and even malicious actors who want to exploit the computational resources of regular users through \textit{cryptojacking} malware. Especially with ready-to-use mining scripts easily provided by service providers (e.g., Coinhive) and untraceable cryptocurrencies (e.g., Monero), cryptojacking malware has become an indispensable tool for attackers. Indeed, the banking industry, major commercial websites, government and military servers (e.g., US Dept. of Defense), online video sharing platforms (e.g., Youtube), gaming platforms (e.g., Nintendo), critical infrastructure resources (e.g., routers), and even recently widely popular remote video conferencing/meeting programs (e.g., Zoom during the Covid-19 pandemic) have all been the victims of powerful cryptojacking malware campaigns. Nonetheless, existing detection methods such as browser extensions that protect users with blacklist methods or antivirus programs with different analysis methods can only provide a partial panacea to this emerging cryptojacking issue as the attackers can easily bypass them by using obfuscation techniques or changing their domains or scripts frequently. Therefore, many studies in the literature proposed cryptojacking malware detection methods using various dynamic/behavioral features. However, the literature lacks a systemic study with a deep understanding of the emerging cryptojacking malware and a comprehensive review of studies in the literature. To fill this gap in the literature, in this SoK paper, we present a systematic overview of cryptojacking malware based on the information obtained from the combination of academic research papers, \ege{two large cryptojacking datasets of samples, and 45} major attack instances. Finally, we also present lessons learned and new research directions to help the research community in this emerging area.
\end{abstract}



\textbf{This paper is accepted in IEEE EuroS\&P 2021 and has been uploaded to arXiv for feedback from stakeholders.}

\section{Introduction} \label{sec:introduction}
Since the day Bitcoin was released in 2009, blockchain-based cryptocurrencies have seen an increasing interest beyond specific communities such as banking and commercial entities. It has become so trivial and ubiquitous to conduct business with cryptocurrencies for any end-user as most financial institutions have already started to support them as a valid monetary system. Today, there are more than 2000 cryptocurrencies in existence. Especially in 2017, the interest for cryptocurrencies peaked with a total market value close to \$1 trillion~\cite{total-cap-2017}. According to a recent Kaspersky report~\cite{GlobalRanks}, 19\% of the world's population have bought some cryptocurrency before. However, buying cryptocurrency is not the only way of investing. Investors can also build mining pools to generate new coins to make a profit. Profitability in mining operations also attracted attackers to this swiftly-emerging ecosystem.

\textit{Cryptojacking} is the act of using the victim's computational power without consent to mine cryptocurrency. This unauthorized mining operation costs extra electricity consumption and decreases the victim host's computational efficiency dramatically. As a result, the attacker transforms that unauthorized computational power into cryptocurrency. 
In the literature, the malware used for this purpose is \ege{known as} \textit{cryptojacking}. 
Especially after the emergence of service providers (e.g., Coinhive~\cite{Coinhive/Authedmine}, CryptoLoot~\cite{crypto-loot}) offering ready-to-use implementations of in-browser mining scripts, attackers can easily reach a large number of users through popular websites.

\vspace{3pt}
\noindent \textbf{In-browser \ege{cryptojacking} examples.} In a major attack, cryptojacking malware was merged with 
Google's advertisement packages on Youtube~\cite{MineYoutube}. The infected ads package compiled by victims' host 
performed unauthorized mining as long as victims stayed at the related page. Youtube and similar media content providers are ideal for the attackers because of their relative trustworthiness, popularity, and average time spent on 
those webpages by the users. In another incident, cryptojacking malware was found in a plugin provided by the UK government~\cite{UKGovMine}. At the time, this plugin was 
in use by several thousands of governmental and non-governmental webpages.
 
\vspace{3pt} 
\noindent \textbf{Cryptojacking examples found on critical servers.} In addition to cryptojacking malware embedded into webpages, cryptojacking malware has also been found in \textit{well-protected governmental and military servers}. The USA Department of Defense discovered cryptojacking malware in their servers during a bug-bounty challenge~\cite{DOD}. The cryptojacking malware found in the DOD servers was created by the famous service provider Coinhive~\cite{Coinhive} and mined 35.4 Monero coin during its existence. Similarly, another governmental case came up from the Russian Nuclear Weapon Research Center~\cite{russian-nuclear}. Several scientists working at this institution were arrested for uploading cryptocurrency miners into the facility servers. Moreover, attackers do not only use the scripts provided by the service providers but also modified the non-malicious, legitimate, open-source cryptominers. For example, 
a cybersecurity company detected an irregular data transmission to a well-known European-based botnet from the corporate network of an Italian bank~\cite{italian-bank}. Further investigation identified that this malware was, in fact, a Bitcoin miner. 

\vspace{3pt}
\noindent \textbf{Cryptojacking examples utilizing advanced techniques.} There have also been many incidents where the attackers used \textit{advanced techniques} to spread cryptojacking malware. For example, in an incident, a known botnet, Vollgar, attacked all MySQL servers in the world~\cite{microsoft-mssql} to take over the admin accounts and inject cryptocurrency miners into those servers. Another recent incident was reported for the Zoom video conferencing program~\cite{ZoomMiner} during the peak of the Covid-19 pandemic, in which 
the attacker(s) merged the main Zoom application and cryptojacking malware and published it via different file-sharing platforms. In other similar incidents, attackers used gaming platforms such as Steam~\cite{GameMiner} and game consoles such as Nintendo Switch~\cite{NintendoSwitch} to embed and distribute cryptojacking malware. Last but not least, in a recent study~\cite{bijmans2019just}, researchers discovered a firmware exploit in Mikrotik routers that were used to embed cryptomining code into every outgoing web connection, where 1.4 million MikroTik routers were exploited.

 
\vspace{3pt}
\noindent \textbf{Challenges of cryptojacking detection.} Given the prevalent emerging nature of the cryptojacking malware, it is vital to detect and prevent unauthorized mining operations from abusing any computing platform's computational resources without the users' consent or permission. However, though it is critical, detecting cryptojacking is challenging because it is different from traditional malware in several ways. First, they abuse their victims' computational power instead of harming or controlling them as in the case of traditional malware. Traditional malware detection and prevention systems are optimized for detecting the harmful behaviors of the malware, \ege{but} cryptojacking malware only uses \ege{computing} resources \abbas{and sends back the calculated hash values to the attacker;} so the malware detection systems commonly consider cryptojacking malware as a heavy application that needs high-performance usage. Second, they can also be used or embedded in legitimate websites, which makes them harder to notice because those websites are often trustworthy, and users do not expect any nonconsensual mining on their computers. \abbas{Third, while in traditional malware attacks, the attacker may ultimately target to exfiltrate sensitive information (i.e., Advanced Persistent Threat (APT)), to make the machine unavailable (i.e., Distributed Denial of Service (DDoS)) or to take control of the victim's machine (i.e., Botnet), in cryptojacking malware attacks, the attacker's goal is to stay stealthy on the system as long as possible since the attack's revenue is directly proportional to the time a cryptojacking \egee{malware} goes undetected.} 
Therefore, attackers use filtering and obfuscation techniques that make their malware harder for detection systems and harder to be noticed by the users. 

\vspace{3pt}
\noindent \textbf{Our contributions.} Due to the seriousness of this emerging threat and the challenges presented above, many cryptojacking studies have been published before. 
However, these studies are either proposing a detection or prevention mechanism against cryptojacking malware or analyzing the cryptojacking threat landscape. And, 
the literature lacks a systemic study covering both different cryptojacking malware types, techniques used by the cryptojacking malware, and a review of the cryptojacking studies in the literature. In this paper, \ege{to fill this gap in the literature, we present a systematic overview of cryptojacking malware based on the information obtained from the combination of 42 
cryptojacking research papers, $\approx26K$ cryptojacking samples with two unique datasets\deleted{compile}, and 45 major attacks instances.} Given the widespread usage of cryptojacking,  
it is important to systematize the cryptojacking malware knowledge for 
the security community 
to accelerate further practical defense solutions against this ever-evolving threat.

\vspace{3pt}
\noindent \textbf{Key takeaways.} In addition to the systematization of cryptojacking knowledge and review of the literature, some of the key takeaways from this study are as follows:

\begin{itemize}[leftmargin=0.10in]

\item \deleted{In both datasets, we observed that even though in-browser cryptojacking attacks have not vanished at all, there is a meaningful decrease in the number of in-browser cryptojacking attacks due to the Coinhive's shutdown concurrent with the actions taken by the platforms. The security reports also observe the same drop} 
\added{Recently, security reports~\cite{IBMreport2020,checkpointreport,JackingIoTReport} spotted}
some increase in the number of cryptojacking attacks targeting more powerful platforms such as cloud servers\added{\cite{DOD, AWS-Hack}}, Docker engines\cite{cloud-docker}, IoT devices on large-scale Kubernetes clusters\added{~\cite{JackingIoTReport}}. 
To hijack and gain initial access\added{~\cite{SocialEng, cve_monero}} to spread the cryptojacking malware, the attackers utilize:
\begin{enumerate}[leftmargin=0.35in]
        \item hardware vulnerabilities \cite{MikrotikNews},
        \item recent CVEs \cite{cve_monero},
        \item poorly configured IoT devices \cite{MiraiBitcoin},
        \item Docker engines and Kubernetes clusters~\cite{cloud-docker} with poor security,
        \item popular DDoS botnets for the side-profit\cite{MiraiBitcoin}.
    \end{enumerate}
\deleted{Despite this change in cryptojacking malware's behavior, t}\added{T}his new trend of cryptojacking malware has not been investigated in detail by researchers.
    
\item We identified several issues in the studies proposing cryptojacking detection mechanisms in literature. First, we found that as the websites containing cryptomining scripts are updated frequently, it is important for the proposed detection studies to report the dataset dates, which is not very common in the studies in the literature. Second, it is important to report if the proposed detection is online or offline, which is missing in most studies.  Third, we also note that the studies in the literature do not measure the overhead on the user side of the proposed solutions, which is critical, especially for browser-based solutions. 

\item \ege{We see 
that 
although} cryptomining could be an alternative funding mechanism for legitimate website owners such as publishers or non-profit organizations, this usage with 
the in-browser cryptomining has diminished due to the keyword-based automatic detection systems.
\end{itemize}

\noindent \textbf{Other surveys.} In the literature, a number of blockchain or Bitcoin-related surveys have been published. However, these surveys only focus on consensus protocols and mining strategies in blockchain~\cite{wang2019survey,lin2017survey,sankar2017survey,wang2018survey}, challenges, security and privacy issues of Bitcoin and blockchain technology~\cite{bonneau2015sok,li2020survey,feng2019survey,jesus2018survey,conti2018survey,10.1145/3316481}, and the implementation of blockchain in different industries~\cite{al2019blockchain} such as IoT
~\cite{dai2019blockchain,panarello2018blockchain}. The closest work to ours is Jayasinghe et al.~\cite{jayasinghe2020survey}, where the authors only present a survey of attack instances of cryptojacking targeting cloud infrastructure. \ege{Hence, this SoK paper is the most comprehensive work focusing on cryptojacking malware made with the observations and analysis of two large datasets. 
}

\vspace{2pt}
\noindent \textbf{Organization.} The rest of this systematization paper is organized as follows: In Section~\ref{sec:background}, we provide the necessary background information on blockchain and cryptocurrency mining. Then, in Section~\ref{sec:methodology}, we explain the methodology we used in this paper. After that, in Section~\ref{sec:cryptojackingmalwaretypes}, we categorize cryptojacking malware types and give their lifecycles. In Section~\ref{MalwareTechniques}, we give broad information about the source of cryptojacking malware, infection methods, victim platform types, target \added{crypto}currencies, \added{detection and prevention methods,} and finally, evasion and obfuscation techniques used by the cryptojacking malware. Section~\ref{sec:literature} presents an overview of the cryptojacking-related studies and their salient features in the literature. Finally, in Section~\ref{sec:research_directions}, we summarize the lessons learned and present some \added{potential} research directions in the domain and conclude the paper in Section~\ref{sec:conclusion}.

\section{Background} \label{sec:background}
In this section, we briefly explain the blockchain concept and cryptocurrency mining process in blockchain networks. Note that cryptocurrency mining is a legitimate operation, and it can be used for profit. To see how cryptojacking malware exploits this process, we first explain how this process works.

\subsection{Blockchain}
Blockchain is a distributed digital ledger technology storing the peer-to-peer (P2P) transactions conducted by the parties in the network in an immutable way. 
Blockchain structure consists of a chain of blocks\abbas{.
As an example, in Bitcoin~\cite{nakamoto2008bitcoin}, each block has two parts: block header and transactions.} A block header consists of the following information: 1) Hash of the previous block, 2) Version, 3) Timestamp, 4) Difficulty target, 5) Nonce, and 6) The root of a Merkle tree. 
By inclusion of the hash of the previous block, every block is mathematically bound to the previous one. This binding makes it impossible to change data from any block in the chain. On the other hand, the second part of each block 
includes a set of individually confirmed transactions.

\subsection{Cryptocurrency Mining} \label{subsec:mining}
 The immutability of a blockchain is provided by a consensus mechanism, which is commonly realized by a "Proof of Work" (PoW) protocol. The immutability of each block and the immutability of the entire blockchain are preserved thanks to the chain of block structure.
 In PoW, some nodes in the network solve a hash puzzle to find a unique hash value and broadcast it to all other nodes in the network. The first node broadcasting the valid hash value is rewarded with a block reward and \ege{collects} transaction fees. A valid hash value is verified according to a difficulty target, i.e., if it satisfies the difficulty target, it is accepted by all other nodes, and the node that found the valid hash value is rewarded. Different PoW implementations usually have different methods for the difficulty target.

The miners try to find a valid hash value by trial-and-error by incrementing the nonce value for every trial. Once a valid hash value is found, the entire block is broadcast to the network, and the block is added to the end of the last block. This process is known as \textit{cryptocurrency mining} (i.e., \textit{cryptomining}), and it is the only way to create new cryptocurrencies. The chance of finding of valid hash value by a miner is directly proportional to the miner's hash power, which is related to the computational power of the underlying hardware. However, more hardware also increases electricity consumption. Therefore, attackers have an incentive to find new ways of increasing computational power without increasing their own electricity consumption.

Following the invention of Bitcoin, many other alternative cryptocurrencies (i.e., altcoins) have emerged and are still emerging. These new cryptocurrencies either claim to address some issues in Bitcoin (i.e., scalability~\cite{kurt2020ln}, privacy) or offer new applications (i.e., smart contracts~\cite{bhargavan2016formal}). 

In the early days of Bitcoin, the mining was performed with the ordinary Central Processing Unit (CPU), and the users 
could easily utilize their regular CPUs for Bitcoin mining. Over time, \ege{Graphical Processing Unit (GPU)-based miners gained significant advantages over CPU miners as GPUs were specifically designed for high computational performance for 
heavy applications.
Later, Field Programmable Gate Array (FPGA) \ege{have} changed the cryptocurrency mining landscape as they were customizable hardware and provided significantly more profit than the CPU or GPU-based mining. Finally, the use of the Application-Specific Integrated Circuit (ASIC) based mining has recently dominated the mining industry as they are specially manufactured and configured for cryptocurrency mining.}

%
\abbas{The alternative} cryptocurrencies also used different hash functions in their blockchain structure, which led to variances in the mining process. For example, Monero~\cite{van2013cryptonote} uses the CryptoNight algorithm as the hash function. CryptoNight is specifically designed for CPU and GPU mining. It uses L3 caches to prevent ASIC miners. With the use \ege{of RandomX\cite{van2013cryptonote} algorithm}, Monero blockchain fully eliminated the ASIC miners and increased the advantage of the CPUs significantly. This feature makes Monero the only major cryptocurrency platform that was designed specifically to favor CPU mining to increase its spread. Moreover, Monero is also known as a private cryptocurrency, and it provides untraceability and unlinkability features through mixers and ring signatures. Monero's both ASIC-miner preventing characteristics and privacy features make it desirable for \egee{attackers}. 

\section{SoK Methodology} \label{sec:methodology}
\vspace{-8pt}
In this section, we explain the sources of information and the methodology used throughout this \ege{SoK} paper. Particularly, we benefit from the papers, recent cryptojacking samples we collected, and publicly known major attack instances.
\vspace{-5pt}
\subsection{Papers}
\vspace{-5pt}
Cryptomining and cryptojacking have recently become popular topics among researchers after the price surge of cryptocurrencies and the release of Coinhive cryptomining script in 2017. For our work, \abbas{we scanned the top computer security conferences (e.g., USENIX) and journals (e.g., IEEE TIFS) given in \cite{google_scholar} as well as the digital libraries (e.g., IEEEXplore, ACM DL) with the keywords such \added{as} cryptojacking, bitcoin, blockchain, etc. and a variety of combinations of these keywords.} In total, we found 43 cryptojacking-related papers in the literature. While one of the papers \abbas{\cite{jayasinghe2020survey}} is a survey paper, the rest are focusing on two separate topics: 1) Cryptojacking detection papers, 2) Cryptojacking analysis papers. We found that there are 15 cryptojacking analysis papers, while there are 27 cryptojacking detection papers in the literature. We further present a literature review of these 42 studies in Section \ref{sec:literature}. Figure \ref{fig:paperdistribution} in Appendix \ref{sec:paper_distribution} shows the distribution of research cryptojacking-related research papers per year. As seen in 
the figure, there is an increasing effort in the academia in the last three years with 
many research papers. Therefore, there is a need for a systematization of this knowledge for cultivating better solutions.
\vspace{-7pt}
\subsection{Samples}
\vspace{-7pt}
Each research paper in the literature focuses only on one aspect of the cryptojacking malware. For a comprehensive understanding of the cryptojacking malware, we also benefited from the real cryptojacking malware samples. \abbas{For this purpose, we \added{collected} \deleted{created} two datasets: 1) \ege{VirusTotal (VT) Dataset} and 2) PublicWWW Dataset. 
\added{The VT dataset consists of 20200 cryptocurrency "miner" samples uploaded to the VT \cite{VirusTotal} and their VT scan reports. On the other hand, we created the PublicWWW dataset using the website source search engine PublicWWW \cite{publicwww}. We found 6269 unique domain containing cryptomining script in their source code.}
\deleted{Particularly, w}\added{W}e used these two datasets for the following purposes and in the noted sections:}

\begin{itemize}
    \item \abbas{To understand the lifecycle of in-browser and host-based cryptojacking (Section \ref{subsec:browsercryptojacking} \& \ref{subsec:hostbased})}
    \item To verify the service provider list given in other studies and as a source of cryptojacking malware (Section \ref{subsec:service_provider})
    \item{To verify the use of \deleted{mobile devices as a victim platform} \added{mobile filtering methods used in the in-browser cryptojacking malware.} (Section \ref{subsubsec:mobile})}
    \item \abbas{To verify that Monero is the main target currency used by cryptojackings (Section \ref{subsubsec:monero})}
    \item \abbas{To find the other cryptocurrencies used by cryptojacking malware (Section \ref{subsec:target})}
    \item \abbas{To verify that the existence of CPU limiting technique for the obfuscation (Section \ref{subsec:cpu-limiting})}
    \item \abbas{To verify and understand the use of code encoding for the obfuscation (Section \ref{encoding})}
\end{itemize}
We note that even though these two datasets were useful for these purposes, they also have some limitations that may affect the findings in this paper. For the VT dataset, we are given $\approx437K$ unique samples (cryptojacking and non-cryptojacking) by the VT \cite{VirusTotal} using our Academic API privileges and their scan reports. Therefore, the VT dataset \lastdeleted{may not be the}\lastadded{is not an} exhaustive list of cryptojacking samples on the VT because API accesses with more extended privileges exist. \lastadded{For example, we observed that $85\%$ of all cryptojacking samples in the VT dataset are from 2018, showing that 2018 samples are over-represented in the VT dataset. Figure \ref{fig:VT_dataset_per_year} in Appendix \ref{subsec:vt_dataset} shows the distribution of all samples by the submission date.} Therefore, any conclusion in terms of the representation of the samples in real-life may have a bias. However, we also note that such a large-scale measurement study is outside the scope of this work. For example, the study in \cite{pastrana2019first} presents such a study with 1.2M malicious cryptocurrency miners collected over a period of twelve years. In our paper, we focused on understanding the behavioral characteristics of cryptojacking malware and reviewing the studies in the literature. 
\deleted{Moreover, w}\deleted{W}e give a more detailed explanation \added{of the datasets, their limitations, and} perform distribution analysis of these datasets in Appendix. Finally, we also published our dataset\added{s}\footnotemark to further accelerate the research in this field.

\footnotetext{\url{https://github.com/sokcryptojacking/SoK}}

 
\subsection{Major Attack Instances}
Our third source of information is the major attack instances that appeared on the 
security reports released by the security companies such as Kaspersky, Trend Micro, Palo Alto Network, \abbas{IBM}, and others, as well as the major security instances that appeared on the news. \abbas{
The major attack instances that appeared on the news may be used to identify unique and interesting cases, while the security reports may shed light on the trends due to the real-time and large-scale reach of the security companies. Particularly, we used these instances in Section \ref{sec:introduction} for motivational purposes; in Section 5 to find out different techniques used by cryptojacking malware; in Section 7 in order to find out potential new trends in the cryptojacking malware attacks. Since 
the collection of these resources may be valuable to other researchers and due to the space limitation here, we also release them in a detailed and organized way together with the blacklists and service providers' documentation links in our dataset link\footnotemark[\value{footnote}].}
Table \ref{fig:attack_distribution} in Appendix \ref{sec:paper_distribution} shows the yearly distribution of the attack instances we used in this paper.


\vspace{-5pt}
\section{Cryptojacking Malware Types} \label{sec:cryptojackingmalwaretypes}
\vspace{-5pt}
Cryptojacking malware, also known as cryptocurrency mining malware, compromises the computational resources of the victim's device (i.e., computers, mobile devices) without the authorization of its user to mine cryptocurrencies and receive rewards. A cryptojacking malware\abbas{'s lifecycle} consists of three main phases: \textit{1) script preparation, 2) script injection,} and \textit{3) \egee{the attack}}. The script preparation and \egee{attack} phases are the same for all cryptojacking malware types. In contrast, the script injection phase is conducted either by injecting the malicious script into the websites or locally embedding the malware into other applications. Based on this, we classify the cryptojacking malware into two categories: \textit{1) In-browser cryptojacking} and \textit{2) Host-based cryptojacking}. In the following sub-sections, we explain the lifecycle of both in-browser and host-based cryptojacking malware. 

\subsection{Type-I: In-browser Cryptojacking}\label{subsec:browsercryptojacking}
The development of web technologies such as JavaScript (JS) and WebAssembly (Wasm) enabled interactive web content, which can access the \ege{several computational resources (e.g., CPU) of the victim's device (e.g., computer or mobile device).} In-browser cryptojacking malware uses these web technologies to create unauthorized access to the victim's system for cryptocurrency mining via \ege{web page interactions on the victim's CPU.} 

\begin{figure}[h]
    \centering
    \framebox{
    \includegraphics[width=.80\columnwidth,trim=0cm 0cm 0cm 0cm]{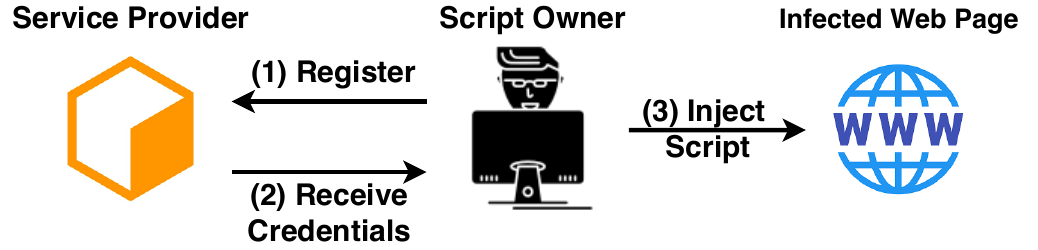}} 
     \caption{Script preparation and injection phases of a in-browser cryptojacking malware.}
    \label{fig:preperation}
\end{figure}

Figure~\ref{fig:preperation} shows the script preparation and injection phases of in-browser cryptojacking malware. The script owner\footnote{We call it script owner rather than an attacker because the script can also be used for legitimate purposes.} first registers (Step 1) and receives its service credentials and ready-to-use mining scripts from the service provider (Step 2). The service provider separates the mining tasks among its users and collects all the revenue from the mining pool later to be shared among its users. After receiving the service credentials, the script owner injects the malicious cryptojacking script into the website's HTML source code (Step 3). We explain this and other cryptojacking infection methods in Section \ref{subsec:infection} in detail.

 \begin{figure*}[t]
    \centering
    \framebox{
    \includegraphics[width=.85\textwidth]{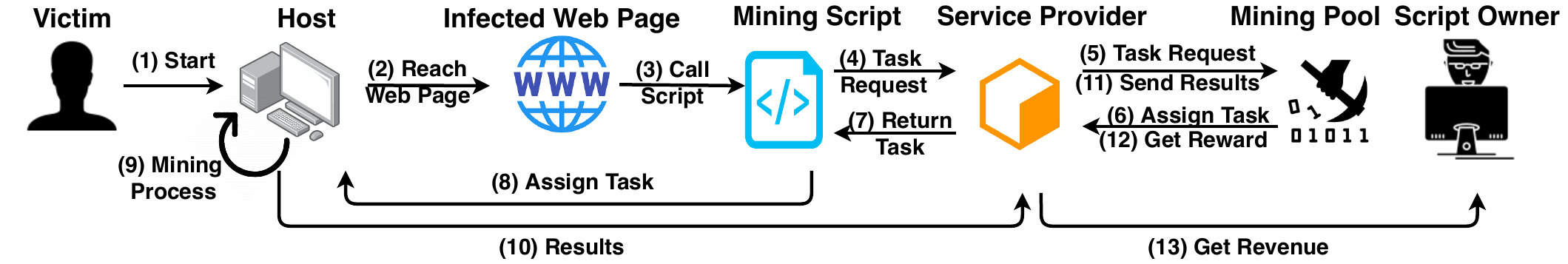}}
    \caption{The lifecycle of a in-browser cryptojacking malware.}
    \label{fig:in-browser-lifecycle}
\end{figure*}

In the \egee{attack} phase, as shown in Figure~\ref{fig:in-browser-lifecycle}, \ege{victims} first \added{reach the} website source code from \ege{their} devices (Step 1,2). The web browser \ege{loads} the website and automatically calls the cryptojacking mining script (Step 3). Once the script is executed, it requests a mining task from the service provider (Step 4). The service provider transfers the task request to the mining pool (Step 5). Then, the mining pool assigns the mining task (Step 6). The service provider returns the task to the mining script (Step 7). The mining script returns this new mining assignment to the victim's computer (Step 8), and the victim's device starts the mining process (Step 9). As long as the mining script and service provider remain online, the script continues the mining process on the victim's computer (Step 9) and then returns the mining results to the service provider (Step 10) directly. The service provider collects all the data from different sources and sends the results to the mining pool (Step 11). Finally, the mining pool sends the reward back to the service provider in the form of a mined currency (Step 12). The script owner receives its share from the service providers using its service credentials after the service provider cuts its service fee. \ege{In this ecosystem, the attackers use\deleted{s} the CPU power of \ege{their} victims, and the \ege{victims} do
not receive any payment nor benefit from any other entity.}
\vspace{-7pt}
\subsection{Type-II: Host-based Cryptojacking} \label{subsec:hostbased}
Host-based cryptojacking is a silent malware that attackers employ to access the victim host's resources and to make it a \ege{zombie} computer for the malware owner. Compared to in-browser cryptojacking malware, host-based malware does not access the victim's computation power through a web script; instead, they need to be installed on the host system. Therefore, \ege{they are generally delivered to the host system through methods such as embedded into third-party applications~\cite{ZoomMiner,utorrent-mining}, using  vulnerabilities~\cite{cve_monero}, or social engineering techniques~\cite{MailMiner}, or as a payload in the drive-by-download technique~\cite{DrivebyMiner}. We explain these methods in more detail in Section~\ref{subsec:infection}.}

\begin{figure}
    \centering
    \includegraphics[width=.85\columnwidth,trim=0cm 0cm 0cm 0cm]{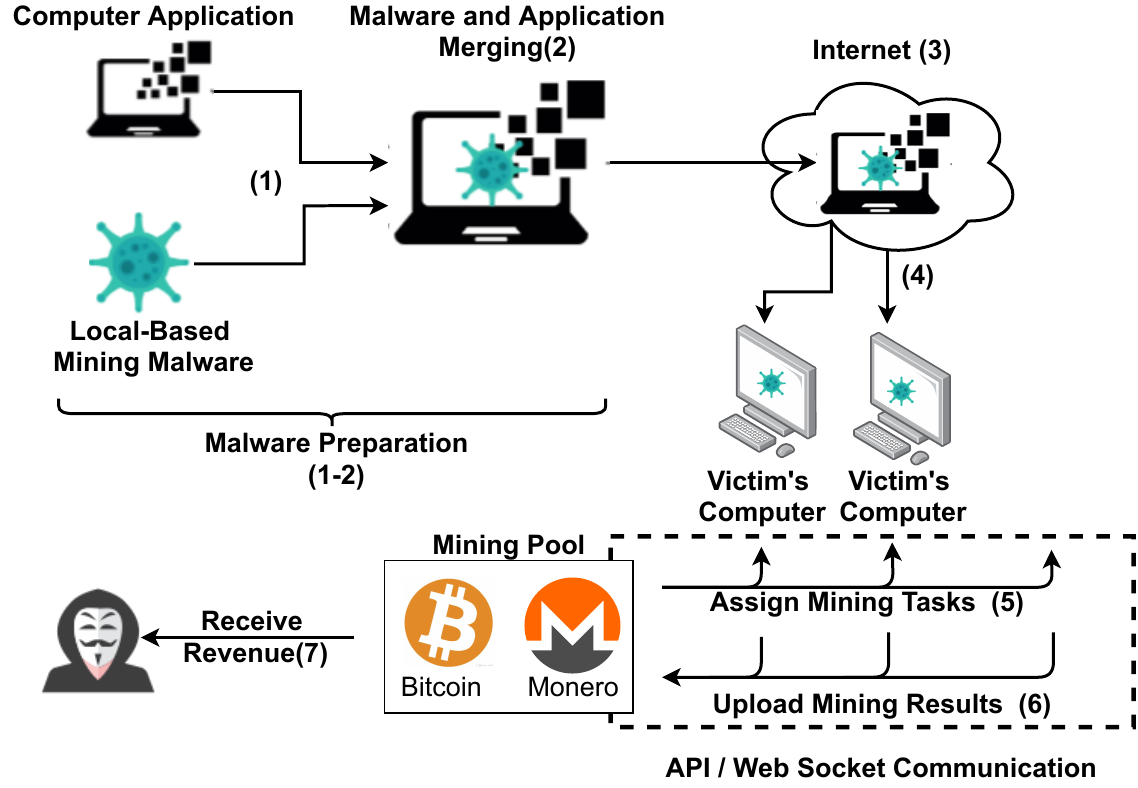}
    \caption{The lifecycle of a host-based cryptojacking malware.}
    \label{fig:host-based cryptojackking malware}
\end{figure}

Figure~\ref{fig:host-based cryptojackking malware} shows the lifecycle of a host-based cryptojacking malware. The script preparation phase starts with the creation of unauthorized cryptocurrency mining malware (1). Then, the attacker merges this malware with a legitimate application to trick the victim (2). After the malware preparation, the malware injection process starts with uploading this malicious application to online data-sharing platforms (e.g., torrent, public clouds) (3). When the victim downloads any of the infected applications and installs them on their host machines (e.g., \ege{ Personal Computer}, IoT device, Server)(4), the malware injection phase of the lifecycle is completed.

In the \ege{attack phase}, the host-based cryptojacking malware is connected to the mining pool via web socket or API and receives the hash puzzle tasks to calculate hash values (5). The calculated hash values are sent back to the mining pool (6). Finally, the attacker receives all of the revenue without any energy consumption (7) and not sharing anything with the victim.

After receiving all its revenue in the form of cryptocurrency from the service provider, the attacker has three options to use its revenue: 1) Converting to fiat currency via exchanges or p2p transactions, 2) Using it as a cryptocurrency for a service~\cite{lee2019cybercriminal}, or 3) Using cryptocurrency mixing services~\cite{goriacheva2018anonymization,bonneau2014mixcoin} to cover its traces. Further end-to-end analysis of the cryptojacking economy/payments is out of this study's scope, and similar studies can be found in the ransomware domain~\cite{huang2018tracking,paquet2019ransomware,conti2018economic,kharraz2015cutting}.

\section{Cryptojacking Malware Techniques}\label{MalwareTechniques}
In this section, we explain the techniques used by cryptojacking malware. Particularly, we articulate on the \ege{following}:

\begin{itemize}
    \item Source of cryptojacking malware
    \item Infection methods
    \item Victim platform types
    \item Target cryptocurrencies
    \item Evasion and obfuscation techniques
\end{itemize}

\vspace{-5pt}
\subsection{Source of Cryptojacking Malware}
\vspace{-5pt}
This sub-section explains whom the scripts are created by and how they are distributed to attackers.

\subsubsection{Service Providers} \label{subsec:service_provider}
The service providers are the leading creators and distributors of cryptojacking scripts. The service providers give every user a unique ID to distinguish them in terms of the hash power.
The service provider generates the script for the user regardless of the user is malicious or not. 
All the user needs to do is copy and paste the script to create a malicious sample for the attack. 

Coinhive~\cite{Coinhive} was the first service provider to offer a ready-to-use in-browser mining script in 2017 to create an alternative income for web site and content owners. Even though the initial idea of Coinhive was to provide an alternative revenue to webpage owners, it rapidly became popular among attackers. \deleted{We also observe this for the samples in the VT dataset. While there are only 272 maliciously labeled cryptominer samples uploaded to VT before 2018, there are 17102 malicious cryptominer samples uploaded in 2018 as shown in Figure \ref{fig:VT_dataset_per_year} in Appendix. This huge increase indicates the effect of Coinhive on the popularity of cryptojacking malware.} During the operation of Coinhive, they were holding a significant share of the total hash rate of Monero. After the sharp decrease in Monero's price~\cite{MoneroPrice}, Coinhive was shut down by their owners in March 2019 due to the business' being no longer profitable.

Some of the alternative service providers which had continued/continuing their operations are Authedmine \cite{Coinhive/Authedmine}, Browsermine \cite{Browsermine}, Coinhave\cite{Coinhave}, Coinimp \cite{Coinimp}, Coin nebula \cite{CoinNebula}, Cryptoloot \cite{crypto-loot}, DeepMiner \cite{DeepMiner}, JSECoin \cite{JSECoin}, Monerise \cite{Monerise}, Nerohut \cite{Nerohut}, Webmine \cite{webmine}, WebminerPool \cite{Webminerpool}, and Webminepool \cite{webminepool}. 
Some of these service providers also came up with several new functionalities, such as offering a user notification method or a GUI for the user to adjust the cryptomining parameters. Note that, we also verified these service providers using the samples in \abbas{the PublicWWW} dataset. \abbas{In order to find the corresponding service providers of each sample, we performed \ege{a keyword} search on the HTML source code of all samples. We found that 5328 samples use\deleted{d} one of these 14 aforementioned service providers, while 941 samples with unknown service providers. Moreover, we also found out that 144 samples \deleted{were}\added{are} using scripts from multiple service providers in their source codes. More details on the PublicWWW dataset can be found in Appendix.}

\subsubsection{Cryptominer Software}
Blockchain networks rely on several network protocols and cryptographic authentication methods. Miners must be part of these protocols and follow the rules provided and developed by the communities. PoW-based cryptocurrencies also have specific rules for their blockchain networks. Due to blockchain technology's public and open nature, the source code of these miners are published by the communities via code sharing and communal development platforms. Attackers can easily obtain and modify these miners and adopt them to perform mining inside their victims' host machines. Moreover, there are also several plug-and-play style mining applications provided by several mining pools. Attackers are also modifying these applications for cryptojacking. For example, XMRig \cite{XMRRigGithub} is a legitimate high-performance Monero miner implementation, and it is open-source. Its signature is found in several highly impactful attacks affecting millions of end devices around the world \cite{XMRRigUnit42,XMRigZeus}, which are also reported by Palo Alto Networks and IBM. \abbas{Moreover, we also found 139 unique samples that \deleted{were}\added{are} \ege{labeled} with the signature of "xmrig" in our VT dataset. 
}

\vspace{-3pt}
\subsection{Infection Methods} \label{subsec:infection}
\vspace{-3pt}
In \abbas{this section}, we explain the \ege{infection} methods used by cryptojacking malware \abbas{in detail}.

\subsubsection{Website Owners} Website owners, who have admin access to the website's servers, may employ in-browser mining scripts to gain extra revenue or provide in exchange of an alternative option to premium content they provide. \ege{Only with this method, webpage owners may receive the revenue of the script in their webpage. }While some website owners inform their visitors about the cryptomining script they employ, some others do not inform their visitors, and this behavior can be considered as crime \cite{UkranianArrested} in several countries.

\subsubsection{Compromised Websites} Attackers may inject their cryptojacking malware into random web pages that have several vulnerabilities. Indeed the name cryptojacking itself is the combination of "cryptomining" and "hijacking." Ruth et al. \cite{ruth2018digging} state that ten different users created 85\% of all Coinhive scripts they found. The owners of these webpages do not have any information about these scripts; additionally, they do not profit from them.
    
Several works claim that the attackers generally use the same ID for all the infected web pages, making them more traceable. For example, the authors in \cite{bijmans2019inadvertently} reveals the cryptojacking campaigns through this method and discover that most of these campaigns utilize the vulnerabilities such as remote code execution vulnerabilities. When we investigated the common instances related to this domain, one security company found cryptojacking malware inside of the Indian government webpages \cite{IndiaHack}, which \ege{affect} all \textit{ap.gov.in} domains and sub-domains.  

\subsubsection{Malicious Ads} Some attackers embed their cryptojacking malware into JavaScript-based ads and distribute them via mining scripts. With this method, the attackers can reach random users without any extra effort. To make this attack, they do not need to infect any webpage or application. \ege{YouTube \cite{MineYoutube} and Google ad \cite{GoogleTag} services were also infected and the users of these websites and their services became the victims of the cryptojacking attacks. The attackers successfully mined Monero with their visitors. The attackers successfully mined Monero with their visitors.} The advantage of this method is that it allows attackers to reach a large number of visitors when it is embedded into popular websites without getting access to the website's servers. 
    
\subsubsection{Malicious Browser Extensions} Browser extensions can also reach the computer's CPU sources and act like cryptojacking malware located into a webpage. These extensions have a major distinctive difference; they can stay online and perform mining as long as the infected browser remains open independent from the websites accessed by the victim. However, major browser operators like Google announced that they would ban all the cryptomining extensions on their platform regardless of their intention as it is mostly being abused in practice \cite{google}.
    
\subsubsection{Third-party \abbas{S}oftware}\label{3rdsoftware} Merging malware with any market application and publishing it via several sharing platforms is a well-known method among the attackers to spread the malware. Attackers modify the cryptominer software to run cryptojacking in the background and merge it with legitimate applications. The attackers tend to use computation-intensive applications (e.g., animation applications, games with high hardware needs, engineering programs) because the use of those applications means that the victims' system has computationally powerful hardware and the application that host-based cryptojacking malware embedded, have access permission to the needed hardware components of the victim's host system.
 
Several major instances have already happened, such as, one attacker merged Zoom \cite{ZoomMiner} video calling application with \ege{a regular} bitcoin miner and distributed it via several sharing platforms. In another attack, the attackers used a popular video game Fortnite to spread the virus \cite{fortnite} to mine Bitcoin. Unlike the in-browser mining, which became popular in 2017, we found the attack instances using this method even in 2013, where the Bitcoin mining script found as part of the game's code itself \cite{evil_game}. 

\subsubsection{Exploited Vulnerabilities} \label{exloitedvulnerabilities} In several cases, attackers exploit several zero-day vulnerabilities that they found in hardware and software. Attackers inject their mining malware into several devices and make them mine cryptocurrency. There are several important instances happened in the last several years. The most remarkable example directly affects 1.4 Million Mikrotik \cite{bijmans2019just} routers globally, and a vulnerability in the hardware operating system causes this instance. The researchers claim that a major percentage of Remote Code Execution (RCE) attacks \cite{RCEattack} aims to locate mining scripts inside the host systems. 

\subsubsection{\ege{Social Engineering Techniques}}
\ege{Social engineering is a commonly used technique among malware attackers to bypass security practices. Similarly, \egee{attackers} also use social engineering attacks to manipulate human psychology and navigate the victims' access or install malicious software on their computers. The researchers have observed that attackers are still using old techniques such as social engineering to install cryptojacking malware on their victims' computers \cite{SocialEng}. }

\subsubsection{\ege{Drive-by Download}}
\ege{
A drive-by download is another technique used by malware attackers to deliver and install malicious files to victims' devices without their knowledge. Victims may face this attack while visiting a web page, opening a pop-up window, or checking an email attachment. In one case \cite{DrivebyMiner}, the attackers used this method to inject their cryptojacking malware into their victims' devices. They exploited shell execution vulnerability to download their cryptojacking malware to victims' computers directly.
}

\subsection{Victim Platform Types}

\subsubsection{Browser}\label{browsercryptojacking}
Browsers are the most commonly used victim platforms as the attackers do not need to deliver any malicious payload to the victim to use the computational resources of the victim. In other words, when the victim reaches the infected webpage, the malware automatically starts mining and do not leave any data behind. The second significant advantage of the browser environment is, thanks to service providers, ready-to-use mining scripts can be applied to any webpage very easily and quickly. The studies in the literature that we also present in Section \ref{sec:literature} mostly focus on in-browser cryptojacking. However, the attackers can access only the CPUs of the victims through the browsers, which makes them infeasible for the currencies allowing ASIC miners such as Bitcoin. Therefore, cryptojacking malware samples utilizing browsers mostly mine Monero or other cryptocurrencies, which allow cryptomining by personal computers on non-ASIC CPU architectures.

\subsubsection{\ege{Personal Computers}}{
\ege{
Personal computers are generally designed to allow end-users to perform their daily tasks. Personal computers are recently modified to overcome high-level computations to allow their users to use heavy-computation applications (e.g., video-games, video rendering applications). Attackers targeting personal computers aim to reach many victims because a limited number of victims would not be profitable. In-browser cryptojacking embedded into popular websites is ideal for this type of cryptojacking attack. In addition, they can also instantiate such an attack through large-scale campaigns. For example, in \cite{PC-vivin}, Cisco researchers document their findings of a two-year campaign delivering XMRig in their payload. They also observed that the malware "makes a minimal effort to hide their actions" and posting the malware "on online forms and social media" to increase the victim pool.
}
}

\subsubsection {\ege{On-premise} Server}
\abbas{On-premise (i.e., in-house) servers are the servers where the data is stored and protected on-site. It is preferred by highly critical organizations such as governmental organizations as it offers greater security and full control over the hardware and data. }
\abbas{However, on-premise servers are also another} victim platform type attacked by the host-based cryptojacking malware samples. Compared to personal computers, \ege{on-premise} servers are more computationally powerful and host numerous services accessed by many connections. This allows attackers to the broader attack surface. \ege{Still}, the attackers have to find a way to deliver and install the cryptomining script on the \ege{on-premise} server to access this platform. In several instances, the attackers used system vulnerabilities \cite{italian-bank}, third party infected software \cite{UKGovMine}, and several social engineering methods \cite{SocialEng} to install cryptojacking malware to the victims' \ege{on-premise} server.


\subsubsection{Cloud Server}
Cryptojacking malware also exploits cloud resources to mine cryptocurrencies. Cloud-based cryptojacking attack is a fast-spreading problem in the last two years, where it became popular, especially after the shutdown of the Coinhive when the attackers were looking for new platforms to infect. \ege{Attackers target several vulnerabilities to hijack victims' cloud servers and locate cryptocurrency miners into their systems.} Clouds servers, especially Infrastructure-as-a-service platforms such as Amazon Web Services (AWS), are being targeted by the attackers because of their:

\begin{itemize}
    \item Virtually infinite resources,
    \item Large attack surface due to server structure,
    \item Malware spreading capabilities,
    \item Reliable Internet connection,
    \item Longer mining/profit period due to host-based capabilities
\end{itemize}

Several instances of this type of cryptojacking malware have been found on \abbas{cloud servers \cite{cloud-cpr-report,AWS-Hack,cloud-docker,watchdog,cve_monero,cloud-Azure}. In these attacks, attackers used different techniques to hijack the cloud server to inject cryptojacking malware. 
For example, in their 2020 annual report, Check Point Research \cite{cloud-cpr-report} observed that attackers integrate the cryptominer to the popular DDoS botnets such as KingMiner targeting Linux and Windows servers for side-profits. In another attacker instance \cite{cloud-docker}, the researchers found an open directory containing malicious files. Further analysis revealed that the file contains a DDoS bot targeting open Docker daemon ports of Docket servers and ultimately installing and running the cryptojacking malware after the execution of its infection chain. In a similar attack instance \cite{cve_monero}, the researchers noted a cryptojacking malware delivered using a CVE exploitation targeting WebLogic servers. Tesla-owned Amazon \cite{AWS-Hack} and the clients of Azure Kubernetes clusters \cite{cloud-Azure} were exposed to cryptojacking attacks due to poorly configured cloud servers.} Indeed, Jayasinghe et al. \cite{jayasinghe2020survey} showed that the count of cryptojacking malware targeting cloud-based infrastructure is increasing every year and affects more prominent domains such as enterprises.



\vspace{-5pt}
\subsubsection{IoT Botnet}
IoT devices generally have small processing powers to perform basic tasks. It is being expected that there will be more than 21.5 billion IoT devices connected to the internet \cite{IoTM} by 2025. Attackers aim to create botnets with the collaboration of thousands of these IoT devices and perform several attacks such as DDoS due to their small processor, limited hardware, low-level security, and weak credentials, which was also exploited in the example of Mirai botnet's DDoS attack \cite{mirai}. Later, IBM researchers also found that the modified version of the Mirai Botnet also started to mine Bitcoin \cite{MiraiBitcoin}. 
Bartino et al.\cite{bertino2017botnets} states that there are several worms in IoT devices that hijacked them for mining purposes, and Ahmad et al.\cite{ahmad2019new} proposes a lightweight IoT cryptojacking detection system to detect any cryptojacking attack that focuses on IoT devices.

\vspace{-5pt}
\subsubsection{Mobile} \label{subsubsec:mobile}
Cryptojacking malware samples targeting mobile devices inject cryptojacking script into their application and list the application in the application markets. Like every other type of cryptojacking attack, the mobile-based cryptojacking samples also have seen a great increase in the number of attacks. Because of this, both Google \cite{google_apps} and Apple \cite{apple} removed the cryptomining applications from their platforms. However, they still exist in alternative markets \cite{dashevskyi2020dissecting}. The study by Dashevskyi et al. \cite{dashevskyi2020dissecting} focuses on Android-based cryptojacking malware.

Moreover, mobile devices are generally not considered powerful enough for cryptocurrency mining because they generally use more restricted hardware and optimized operating systems (e.g., iOS and Android). Besides, the cryptocurrency mining process consumes extra battery and processing power, which may cause \ege{hardware problems such as overheating and apps to freeze or crash} on mobile devices. Due to these reasons, cryptojacking attacks on mobile devices are not preferred by attackers, and they generally apply a mobile filtering method to opt-out mobile devices.

\begin{figure}[h]
    \centering
    \includegraphics[width=\columnwidth,trim=0cm 1cm 0cm 0cm]{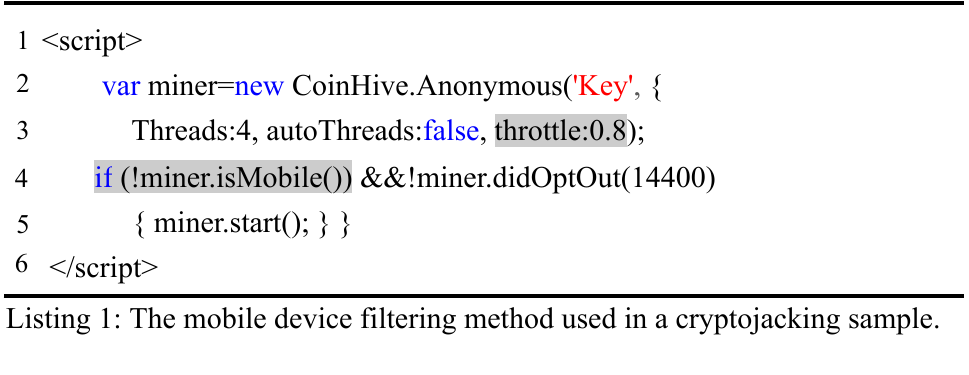}
    \label{listing:mobile_filtering}
    \vspace{-10pt}
\end{figure}


Listing~1 is a recent cryptojacking sample with the mobile device filtering method found in a sample in our dataset. In line 4, the script automatically calls a mobile device detection function and starts the cryptocurrency mining process only if it is not a mobile device.

\vspace{-3pt}

\subsection{Target Cryptocurrencies} \label{subsec:target}
\vspace{-3pt}
In this section, we give brief information about the most preferred cryptocurrencies by the \egee{attackers}.

\subsubsection{Monero} \label{subsubsec:monero}
Monero has several advantages over other cryptocurrencies, making it favorable to \egee{attackers}. \ege{First of all, Monero successfully implements and modifies the RandomX mining algorithm and CryptoNight hashing algorithm to prevent ASIC miners and give a competitive advantage to the CPU miners over GPUs via L3 caches \cite{takahashi2018sha256d}. The Monero community aims to keep their network decentralized and allows even small miners to mine Monero.} As in-browser cryptojacking malware can only access the CPUs of the personal computers through the browsers, \abbas{Monero is ideal as a target cryptocurrency instead of other cryptocurrencies that are mined dominantly by other computationally more powerful ASIC and GPU miners.}
Second, Monero provides anonymity features through cryptographic ring signatures~\cite{bender2006ring,moser2018empirical}, which makes the attackers untraceable. \ege{Thanks to these features of the Monero, attackers tend to mine Monero with their in-browser cryptojacking malware.}

\abbas{When we analyze the samples' cryptomining scripts in the PublicWWW dataset and their service providers' documentation, we found that all 
eleven service providers except Browsermine, CoinNebula, JSEcoin either use Monero or have the option to choose Monero in their scripts as a target cryptocurrency. This shows to 91\% of the samples in the PublicWWW dataset use Monero to mine.}

\subsubsection{Bitcoin}
In recent years, Bitcoin mining has seen enormous attention, which led to a dramatic increase in the difficulty target. \ege{ASIC and FPGA miners are the main reason behind this dramatic increase because the mining structure of the Bitcoin allows to build and use of specified mining hardware which is much more powerful and profitable than the CPUs and GPUs. The increase in difficulty target and disadvantages of CPU made the CPU mining infeasible and not profitable. Therefore, attackers who perform in-browser cryptojacking attack do
not prefer Bitcoin mining. \abbas{We also see that none of \ege{the service} providers of the in-browser cryptojacking samples in our PublicWWW dataset supports Bitcoin mining.} However, host-based cryptojacking malware can reach all the components of the victims' computer system and make Bitcoin mining on GPU and other high-performance computational resources of the computers. \abbas{We also observe this in our VT dataset. We performed a keyword search for "bitcoin" on the AV labels of 20200 samples of both in-browser and host-based cryptojacking malware. We found that 7111 of 20200 samples are marked with a label containing the keyword "bitcoin". Even though this does not show that those samples are absolutely using bitcoin as a target cryptocurrency, but it is a potential indicator for the host-based cryptojacking samples mining Bitcoin based on the assumption of AV vendors are \ege{labeling} the correct currency for the AV labels.
}
}

\subsubsection{Other Cryptocurrencies} Cryptojacking is attractive for attackers as cryptomining can be parallelized among many victims. Therefore, it is possible for cryptocurrencies to allow distributed cryptomining. Both Monero and Bitcoin use PoW as a consensus method. However, instead of PoW, other cryptocurrencies utilize different consensus models such as Proof of Stake~\cite{vasin2014blackcoin}, \ege{and} Proof of Masternode~\cite{duffield2014transaction}. Most of these new consensus models do not depend on distributed power-based mining algorithms; therefore, cryptojacking is not an option for those currencies.
For the cryptocurrencies that can be mined distributively~\cite{darabian2020detecting}, the mining pools provide collective mining services to their participants. Other cryptocurrencies that are preferred by \egee{attackers} are Bitcoin Cash~\cite{bitcoincash}, Litecoin~\cite{Litecoin}, \ege{and Ethereum~\cite{buterin2014next}}.

There are also several cryptocurrencies developed specifically for in-browser cryptomining activities. JSEcoin~\cite{JSECoin} is an example of them and offers also transparency. Other cryptocurrencies created for this purpose are BrowsermineCoin\cite{Browsermine}, Uplexa\cite{Uplexa}, Sumocoin\cite{Sumokoin}, and Electroneum\cite{Electroneum}.  

\vspace{-5pt}
\subsection{\abbas{Detection and Prevention Methods}} \label{subsec:detection_methods}
\vspace{-5pt}

\ege{In the traditional malware detection literature, there are two main analysis methods: 1) static~\cite{nath2014static} and 2) dynamic~\cite{willems2007toward}. Both analysis methods have several pros and cons in terms of accuracy and usability.} 
\begin{itemize}
    \item \textit{Static Analysis:} \ege{Static analysis is a widely used method to examine the application without executing it. Static analysis tools generally seek specific keywords, malware signatures, and hash values. In the cryptojacking domain, mining-blocking browser extensions~\cite{NoCoin,MinerBlock} work
    in this way, i.e., any domain given in the pre-determined blacklist is blocked.
    However, due to the fix, pre-configured 
    nature of the static detection methods, these implementations 
    are usually easy to circumvent.
    }
    \item \textit{Dynamic Analysis:} 
    \ege{In dynamic analysis, the malware sample is executed in a controlled environment, and its behavioral features are recorded for further analysis and detection. Malware analyzers generally use automated~\cite{acar2019enterprise} or non-automated sandboxes~\cite{willems2007toward} to run the code and observe the malware's behavior. In the literature, 24 machine learning-based proposed detection mechanisms use dynamic analysis to detect cryptojacking malware.  These studies use various datasets, features, classification algorithms and some of them works for both in-browser and host-based cryptojacking malware. We explain these studies in Section~\ref{sec:detection}.
    }
\end{itemize} 
\abbas{As the execution of in-browser cryptojacking malware depends on running the JavaScript code, another way to stop it is to disable the use of JavaScript, but this would also decrease the usability of the browser significantly. Finally, there are antivirus programs with the cryptojacking detection capability~\cite{malwarebytes,AvastAntiMAlware}. However, their detection algorithms are proprietary. 
}


\subsection{Evasion and Obfuscation Techniques} \label{obfuscation}

The purpose of the cryptojacking malware is to exploit the resources of the victim as long as possible; therefore, staying on the system without being detected is of paramount importance. For this purpose, they utilize several obfuscation methods.

\subsubsection{CPU \abbas{L}imiting} \label{subsec:cpu-limiting}

High CPU utilization is still the most important common point of all kinds of cryptojacking malware because CPU usage is the main requirement of the cryptocurrency mining process. Therefore, CPU limiting is a highly preferred method by the attackers to obfuscate the mining script. With this method, the script owners can bypass the high CPU usage-based detection systems and avoid being put on the blacklist. Moreover, the CPU limiting is also used by legitimate website owners performing cryptocurrency mining as an alternative revenue because it provides a better user experience. \abbas{Line 3 in} Listing~1 shows an example of a CPU limiting method used by a cryptojacking malware, where the attacker sets \abbas{throttle to $0.8$}
, e.g., the attacker wants to use only $20\%$ of the CPU load for cryptocurrency mining. \abbas{In our PublicWWW dataset, we searched for the keyword "throttle: 0.9" and we found that 1384 samples out of all 6269 cryptomining scripts set the throttle to $90\%$, which shows that CPU is limiting is a very common practice among the in-browser cryptomining scripts.}

    \label{listing:CpuLimiting}


\subsubsection{\ege{Hidden Library Calls}} \label{hidlib}
Library calling~\cite{JSLibCall} is a well-known technique used by programmers to make the code more efficient, systematic, and readable. However, it can also be used by the attackers to obfuscate their scripts. Particularly, in order to hide the mining code from the detection methods, the attackers create new scripts that do not have specific keywords. The malicious part of the script is moved to an external library, which is called during the script's execution, and only the code snippet to call this library is included in the main code. 


\subsubsection{\ege{Code Encoding}}\label{encoding}

Encoding the malware source code with several encoding algorithms \ege{provides} invisibility against keyword-based static analysis detection methods such as blacklists. This method transforms the text data into another form, such as Base64, and after this process, the data can only be read by the computers. Some examples of this we found in our \abbas{PublicWWW} dataset are the cryptomining scripts provided by the service providers Authedmine \cite{Coinhive} and Cryptoloot \cite{crypto-loot}. 

\subsubsection{\ege{Binary Obfuscation}}\label{binary}
\abbas{Similar to code encoding technique,} binary obfuscation is a practice among malware authors to hide malicious code from standard string matching algorithms and make it harder to recover by the sandboxes and other dynamic malware detection methods. \abbas{However, they differ in the cryptojacking type that is used to \ege{hiding,} i.e., binary obfuscation is used by the host-based cryptojacking malware while code encoding is used by the in-browser cryptojacking malware.}
\abbas{For binary obfuscation, a}ttackers generally use well-known packers such as UPX. The authors of \cite{pastrana2019first} observe that 30\% of 1.2M binary cryptojacking malware samples are obfuscated, which shows that it is a common practice among the cryptojacking malware attackers, too. 

\section{Literature Review} \label{sec:literature}
The surge of cryptojacking malware, especially after 2017, also \ege{drew} the attention of academia and resulted in many publications. We found these studies focus on three topics: 1) Cryptojacking detection studies, 2) Cryptojacking prevention studies, and  3) Cryptojacking analysis studies.
Among 42 academic research papers, we found that 15 of them focus on the experimental analysis of the cryptojacking dataset. At the same time, 3 of them proposes a method for the detection and prevention of cryptojacking malware together, and 24 of them proposes only a method for the detection of the cryptojacking malware. In the next sub-sections, we give a review of these studies.

\subsection{Cryptojacking Detection Studies}\label{sec:detection}

\begin{table*}
\caption{Cryptojacking malware detection mechanisms in the literature.}
\label{tab:detections}
\resizebox{\textwidth}{!}{%
\begin{threeparttable}
\renewcommand{\arraystretch}{3.3}
\begin{tabular}{ccccccc}
\hline
  
\multicolumn{1}{|c|}{\textbf{Ref}} &
  \multicolumn{1}{c|}{\textbf{Dataset}} &
  \multicolumn{1}{c|}{\textbf{\ege{Type}}} &
  \multicolumn{1}{c|}{\textbf{Method}} &
  \multicolumn{1}{c|}{\textbf{Features}} &
  \multicolumn{1}{c|}{\textbf{Classifier}} &
  \multicolumn{1}{c|}{\textbf{Performance}} \\ \hline \hline
  
  \multicolumn{1}{|c|}{Rüth et al.~\cite{ruth2018digging}} &
  \multicolumn{1}{c|}{\makecell{Three largest TLDs\\ Alexa: 1M}} &
  \multicolumn{1}{c|}{\ege{In-browser}} &
  \multicolumn{1}{c|}{Static} &
  \multicolumn{1}{c|}{\makecell{Wasm signatures}} &
  \multicolumn{1}{c|}{SRSE} &
  \multicolumn{1}{c|}{\makecell{N/A}} \\ \hline
  
    \multicolumn{1}{|c|}{Minesweeper~\cite{konoth2018minesweeper}} &
  \multicolumn{1}{c|}{\makecell{Alexa 1 Million}} &
  \multicolumn{1}{c|}{\ege{In-browser}} &
  \multicolumn{1}{c|}{Static} &
  \multicolumn{1}{c|}{\makecell{Wasm code \\CPU cache events}} &
  \multicolumn{1}{c|}{Matching} &
  \multicolumn{1}{c|}{N/A} \\ \hline

\multicolumn{1}{|c|}{RAPID~\cite{rodriguez2018rapid}} &
  \multicolumn{1}{c|}{Alexa: 330.500} &
  \multicolumn{1}{c|}{\ege{In-browser}} &
  \multicolumn{1}{c|}{Dynamic} &
  \multicolumn{1}{c|}{\makecell{Resource consumption\\ (memory, network, \\and processor) and\\ JavaScript API Events}} &
  \multicolumn{1}{c|}{\makecell{SVM}} &
  \multicolumn{1}{c|}{\makecell{Benign (best): \\
  Precision: 99.99\% \\ Recall: 99.99\%\\ F1: up to 99.99\% \\ Mining (best):\\Precision: 96.54\% \\ Recall: 95.48\%\\ F1: up to 96.0\%  \\ }} \\ \hline

  
\multicolumn{1}{|c|}{Mu\~{n}oz et al.~\cite{i2019detecting}} &
  \multicolumn{1}{c|}{\makecell{Network traffic of six\\ cryptocurrencies using\\ Stratum protocol}} &
  \multicolumn{1}{c|}{\ege{In-browser}} &
  \multicolumn{1}{c|}{Dynamic} &
  \multicolumn{1}{c|}{\makecell{Metadata of inbound\\and outbound network traffic}} &
  \multicolumn{1}{c|}{DT} &
    \multicolumn{1}{c|}{\makecell{Best:\\ Accuracy: 99.9\%\\ Precision: 98.2\%\\ Recall: 90.7\%}} \\\hline
  
\multicolumn{1}{|c|}{CapJack~\cite{ning2019capjack}} &
  \multicolumn{1}{c|}{\makecell{Five user applications\\ and  a Coinhive miner}} &
  \multicolumn{1}{c|}{\ege{In-browser}} &
  \multicolumn{1}{c|}{Dynamic} &
  \multicolumn{1}{c|}{\makecell{CPU utilization,\\Memmory,\\ Disk read/write rate,\\Network interface)}} &
  \multicolumn{1}{c|}{CNN} &
  \multicolumn{1}{c|}{\makecell{87\% Instant\\ 99\% After 11 seconds\\ 98\% Mobile Single \\ 86\% Mobile Cross\\ 97\% AWS single\\ 89\% AWS Cross}} \\ \hline
  
\multicolumn{1}{|c|}{OUTGUARD~\cite{kharraz2019outguard}} &
  \multicolumn{1}{c|}{Alexa 1M and 600K} &
  \multicolumn{1}{c|}{\ege{In-browser}} &
  \multicolumn{1}{c|}{Dynamic} &
  \multicolumn{1}{c|}{\makecell{Js Execution Time,\\ JS compilation Time,\\ Garbage Collector,\\Iframe resource loads,\\CPU usage}} &
  \multicolumn{1}{c|}{\makecell{SVM, RF}} &
  \multicolumn{1}{c|}{\makecell{SVM (best):\\ TPR: 97.9\%\\ FPR: 1.1\%}} \\ \hline
  
  
\multicolumn{1}{|c|}{CoinSpy~\cite{keltonbrowser}} &
  \multicolumn{1}{c|}{\makecell{100k websites from\\ Alexa 1M and 50\\ manipulated cryptojacking\\ websites}} &
  \multicolumn{1}{c|}{\ege{In-browser}} &
  \multicolumn{1}{c|}{Dynamic} &
  \multicolumn{1}{c|}{\makecell{CPU,\\ Memory, \\Network behaviors}} &
  \multicolumn{1}{c|}{CNN} &
  \multicolumn{1}{c|}{Accuracy:97\%} \\ \hline
  
\multicolumn{1}{|c|}{MineCap~\cite{neto2020minecap}} &
  \multicolumn{1}{c|}{\makecell{The network traffic captured\\ from two mining and\\ streaming applications}} &
  \multicolumn{1}{c|}{\ege{In-browser}} &
  \multicolumn{1}{c|}{Dynamic} &
  \multicolumn{1}{c|}{Network packages} &
  \multicolumn{1}{c|}{IL} &
  \multicolumn{1}{c|}{\makecell{Accuracy: 98\%,\\Precision: 99\%,\\ Recall: 97\%\\Specifity: 99.9\%}} \\ \hline
  
\multicolumn{1}{|c|}{CMTracker~\cite{hong2018you}} &
  \multicolumn{1}{c|}{Alexa 100k} &
  \multicolumn{1}{c|}{\ege{In-browser}} &
  \multicolumn{1}{c|}{Dynamic} &
  \multicolumn{1}{c|}{\makecell{Hash and Stuck\\ based profilers}} &
  \multicolumn{1}{c|}{Thr-based} &
  \multicolumn{1}{c|}{100\% TPR} \\ \hline
  
\multicolumn{1}{|c|}{Musch et al.~\cite{musch2019thieves}} &
  \multicolumn{1}{c|}{Alexa 1M} &
  \multicolumn{1}{c|}{\ege{In-browser}} &
  \multicolumn{1}{c|}{Dynamic} &
  \multicolumn{1}{c|}{\makecell{CPU usage}} &
  \multicolumn{1}{c|}{MA} &
  \multicolumn{1}{c|}{N/A} \\ \hline
  
\multicolumn{1}{|c|}{Tahir et al.~\cite{tahir2019browsers}} &
  \multicolumn{1}{c|}{\makecell{Manually created 320 non-mining\\ and 100 mining websites}} &
  \multicolumn{1}{c|}{\ege{In-browser}} &
  \multicolumn{1}{c|}{Dynamic} &
  \multicolumn{1}{c|}{HPC values} &
  \multicolumn{1}{c|}{RF} & 
  \multicolumn{1}{c|}{\makecell{Accuracy: 99.35\%,\\Precision: 100\%,\\Recall: 98\%,\\AUC: 99\%}} \\ \hline
  
\multicolumn{1}{|c|}{SEISMIC~\cite{wang2018seismic}} &
  \multicolumn{1}{c|}{\makecell{500 webpages randomly selected \\from Alexa top 50K}} &
  \multicolumn{1}{c|}{\ege{In-browser}} &
  \multicolumn{1}{c|}{Dynamic} &
  \multicolumn{1}{c|}{\makecell{Wasm instructions}} &
  \multicolumn{1}{c|}{Matching} &
  \multicolumn{1}{c|}{F1: 98\%} \\ \hline
  
  \multicolumn{1}{|c|}{MineThrottle~\cite{bian2020minethrottle}} &
  \multicolumn{1}{c|}{\makecell{Alexa 1M}} &  
  \multicolumn{1}{c|}{\ege{In-browser}} &
  \multicolumn{1}{c|}{Dynamic} &
  \multicolumn{1}{c|}{\makecell{Block-level features \\ CPU usage}} &
  \multicolumn{1}{c|}{Matching} &
  \multicolumn{1}{c|}{FNR: 1.83\% } \\ \hline
  
  \multicolumn{1}{|c|}{Coinpolice~\cite{petrov2020coinpolice}} &
  \multicolumn{1}{c|}{\makecell{47k samples}} &  
  \multicolumn{1}{c|}{\ege{In-browser}} &
  \multicolumn{1}{c|}{Dynamic} &
  \multicolumn{1}{c|}{\makecell{CPU usage, HPC,\\ JS/WASM execution time and features, \\ Throttling-independent timeseries }} &
  \multicolumn{1}{c|}{CNN} &
  \multicolumn{1}{c|}{TPR:97.8 \% FPR: 0.74\% } \\ \hline

  \multicolumn{1}{|c|}{Carlin et al.~\cite{carlin2018detecting}} &
  \multicolumn{1}{c|}{\makecell{Captured Opcode trace\\ packets Virusshare (296 Samples)}} &  
  \multicolumn{1}{c|}{\ege{In-browser}} &
  \multicolumn{1}{c|}{Dynamic} &
  \multicolumn{1}{c|}{\makecell{Opcodes}} &
  \multicolumn{1}{c|}{RF} &
  \multicolumn{1}{c|}{\makecell{TPR: 99.2 \% FPR: 0.9,\\ Precision: 99.2 Recall: 99.2}} \\ \hline
    
  \multicolumn{1}{|c|}{Liu et al.~\cite{liu2018novel}} &
  \multicolumn{1}{c|}{\makecell{1159 samples collected from\\ browsers' memory snapshot}} &
  \multicolumn{1}{c|}{\ege{In-browser}} &
  \multicolumn{1}{c|}{Dynamic} &
  \multicolumn{1}{c|}{\makecell{Heap snapshots\\ Stack Features}} &
  \multicolumn{1}{c|}{RNN} &
  \multicolumn{1}{c|}{\makecell{Precision: ~95, \\ Recall: ~93}} \\ \hline
  
  \multicolumn{1}{|c|}{Rauchberger et al.~\cite{rauchberger2018other}} &
  \multicolumn{1}{c|}{\makecell{ Alexa: 1M}} &
  \multicolumn{1}{c|}{\ege{In-browser}} &
  \multicolumn{1}{c|}{Dynamic} &
  \multicolumn{1}{c|}{\makecell{Web socket traffic}} &
  \multicolumn{1}{c|}{Matching} &
  \multicolumn{1}{c|}{\makecell{N/A}} \\ \hline 
  
  \multicolumn{1}{|c|}{Caprolu et al.~\cite{caprolu2019crypto}} &
  \multicolumn{1}{c|}{\makecell{N/A}} &
  \multicolumn{1}{c|}{\ege{In-browser}} &
  \multicolumn{1}{c|}{Dynamic} &
  \multicolumn{1}{c|}{\makecell{Network traffic}} &
  \multicolumn{1}{c|}{RF,KFCV} &
  \multicolumn{1}{c|}{\makecell{TPR=92\% ,FPR=0.8\%}} \\ \hline
  
  \multicolumn{1}{|c|}{MINOS~\cite{naseemminos}} &
  \multicolumn{1}{c|}{\makecell{WASM Samples collected via \\ PublicWWW}} &  
  \multicolumn{1}{c|}{In-browser} &
  \multicolumn{1}{c|}{Dynamic} &
  \multicolumn{1}{c|}{\makecell{Image frames of \\malicious samples}} &
  \multicolumn{1}{c|}{CNN} &
  \multicolumn{1}{c|}{Accuracy: 98.97\%} \\ \hline

  \multicolumn{1}{|c|}{Yulianto et al.~\cite{yulianto2019mitigation}} &
  \multicolumn{1}{c|}{\makecell{PublicWWW and Blacklists}} &  
  \multicolumn{1}{c|}{\ege{In-browser}} &
  \multicolumn{1}{c|}{Static and Dynamic} &
  \multicolumn{1}{c|}{\makecell{CPU usage}} &
  \multicolumn{1}{c|}{Matching} &
  \multicolumn{1}{c|}{TPR:100\% } \\ \hline
    
  \multicolumn{1}{|c|}{CMBlock~\cite{razali2019cmblock}} &
  \multicolumn{1}{c|}{\makecell{In-browser cryptojacking samples}} &  
  \multicolumn{1}{c|}{\ege{In-browser}} &
  \multicolumn{1}{c|}{Static and Dynamic} &
  \multicolumn{1}{c|}{\makecell{Blacklists \\ Behaviour }} &
  \multicolumn{1}{c|}{N/A} &
  \multicolumn{1}{c|}{N/A} \\ \hline
  
  \multicolumn{1}{|c|}{Gangwal et al.~\cite{conti2019detecting}} &
  \multicolumn{1}{c|}{ \makecell{Combination daily user \\tasks and miners\tnote{1}}} &
  \multicolumn{1}{c|}{\ege{Host-based} and \ege{In-browser}} &
  \multicolumn{1}{c|}{Dynamic} &
  \multicolumn{1}{c|}{\makecell{hardware events (e.g., branch-misses),\\ software events (e.g., page-faults)\\ hardware cache events (e.g., cache-misses)}} &
  \multicolumn{1}{c|}{\makecell{RF, SVM}} &
  \multicolumn{1}{c|}{\makecell{Recall: 97.84\%\\ Precision: 99.7\%\\ Accuracy:98.7\%}} \\ \hline
  
  \multicolumn{1}{|c|}{Lachtar et al.~\cite{lachtar2020cross}} &
  \multicolumn{1}{c|}{\makecell{N/A}} &  
  \multicolumn{1}{c|}{\ege{Host-based} and \ege{In-browser}} &
  \multicolumn{1}{c|}{Dynamic} &
  \multicolumn{1}{c|}{\makecell{CPU instructions }} &
  \multicolumn{1}{c|}{Matching} &
  \multicolumn{1}{c|}{TPR:100 \% FPR: < 2\% } \\ \hline
  
  \multicolumn{1}{|c|}{Tanana et al.~\cite{tanana2020behavior}} &
  \multicolumn{1}{c|}{\makecell{40
\ege{In-browser} and \\ 10 executable-type cryptojacking}} &  
  \multicolumn{1}{c|}{\ege{Host-based} and \ege{In-browser}} &
  \multicolumn{1}{c|}{Dynamic} &
  \multicolumn{1}{c|}{\makecell{CPU utilization share \\ RAM usage}} &
  \multicolumn{1}{c|}{N/A} &
  \multicolumn{1}{c|}{TPR: 81\%} \\ \hline
  
  \multicolumn{1}{|c|}{Ahmad et al.~\cite{ahmad2019new}} &
  \multicolumn{1}{c|}{\makecell{Mixture of Benign \\ and Malicious Network Packages}} &  
  \multicolumn{1}{c|}{\ege{Host-based} and \ege{In-browser}} &
  \multicolumn{1}{c|}{Dynamic} &
  \multicolumn{1}{c|}{\makecell{Network traffic}} &
  \multicolumn{1}{c|}{DCA} &
  \multicolumn{1}{c|}{N/A} \\ \hline
  
  \multicolumn{1}{|c|}{DeCrypto Pro~\cite{mani2020decrypto}} &
  \multicolumn{1}{c|}{\makecell{\~1200 samples}} &  
  \multicolumn{1}{c|}{\ege{Host-based} and \ege{In-browser}} &
  \multicolumn{1}{c|}{Dynamic} &
  \multicolumn{1}{c|}{\makecell{HPC, CPU usage}} &
  \multicolumn{1}{c|}{k-NN, RF, LSTM} &
  \multicolumn{1}{c|}{\makecell{FPR: 2.5,\\ Precision: 96, Recall: 97}} \\ \hline
  
  \multicolumn{1}{|c|}{Darabian et al.~\cite{darabian2020detecting}} &
  \multicolumn{1}{c|}{\makecell{1500 active cryptomining\\ collected from Virustotal in 2018}} &
  \multicolumn{1}{c|}{\ege{Host-based}} &
  \multicolumn{1}{c|}{Static and Dynamic} &
  \multicolumn{1}{c|}{\makecell{System calls,\\opcode sequences}} &
  \multicolumn{1}{c|}{\makecell{RNN, CNN}} &
  \multicolumn{1}{c|}{\makecell{System calls (best):\\ LSTM: Accuracy:99\%\\ F1: 98\% MCC: 98\% \\ FPR:0.6\%} }\\ \hline
  
  \multicolumn{1}{|c|}{Crypto-Aegis~\cite{caprolu2019crypto}} &
  \multicolumn{1}{c|}{\makecell{Network traffic of 3 legitimate mining \\scripts and 3 daily user applications}} &  
  \multicolumn{1}{c|}{\ege{Host-based}} &
  \multicolumn{1}{c|}{Dynamic} &
  \multicolumn{1}{c|}{\makecell{Packet sizes\\Interarrival times}} &
  \multicolumn{1}{c|}{RF} &
  \multicolumn{1}{c|}{TPR:80-84\% FPR: 0.9 - 1.2\%} \\ \hline

\end{tabular}%
  \begin{tablenotes}
    \item[1] The dataset was not available as of writing this paper (November 1, 2020).
    \item[2] Support Vector Machine: SVM, Random Forest: RF, Decision Tree: DT, Convolutional Neural Network: CNN, Recurrent Neural Network: RNN, Incremental Learning: IL, Threshold-based: Thr-based, Manual Analysis: MA, Dendritic Cell Algorithm: DCA, k-Nearest Neighbors: k-NN, Light-weight machine learning models: LSTM, Symantec RuleSpace Engine:SRSE, k-Fold Cross Validation:KFCV
  \end{tablenotes}
\end{threeparttable}
}
\end{table*}

\abbas{In this section, we survey the cryptojacking malware detection studies.} Table~\ref{tab:detections} shows the list of the proposed cryptojacking detection mechanisms in the literature. The following sub-section gives a detailed overview of the dataset, platform, analysis method, features, and classifiers used in these detection mechanisms. 

\subsubsection{Dataset} A dataset is generally used to evaluate the effectiveness of the proposed detection method. Several datasets are commonly used in the cryptojacking malware detection literature. The most common one is Alexa top webpages~\cite{rodriguez2018rapid,kharraz2019outguard,keltonbrowser,hong2018you,musch2019thieves,wang2018seismic}. Alexa sorts the most visited websites on the Internet; however, it does not provide the source code for these websites. Therefore, these studies also used Chrome Debugging Protocol to instrument the browser and collect the necessary information from the websites, except the study~\cite{wang2018seismic}, which works with a limited number (500) of websites.
Moreover, the study in~\cite{kharraz2019outguard} also used known and frequently updated blacklists~\cite{NoCoin,MinerBlock,CoinBlockerLists} to build a ground truth for their training dataset, and then they performed an analysis using Alexa top 1M websites. In addition to the Alexa top websites, the study in~\cite{darabian2020detecting} used a cryptojacking dataset obtained from VirusTotal. They collected 1500 active Windows Portable Executable (PE32) cryptocurrency mining malware samples registered in 2018 and used the Cuckoo Sandbox~\cite{guarnieri2012cuckoo} to obtain detailed behavioral reports on those samples. Furthermore, the studies in~~\cite{conti2019detecting,i2019detecting,ning2019capjack,neto2020minecap} performed their analysis by installing the legitimate mining scripts, and the studies in~\cite{keltonbrowser,tahir2019browsers} manually injected miners to the websites to test their detection mechanisms.

\subsubsection{Platform} Most of the cryptojacking detection mechanisms in the literature~\cite{rodriguez2018rapid,conti2019detecting,ning2019capjack,kharraz2019outguard,keltonbrowser,neto2020minecap,hong2018you,musch2018web,tahir2019browsers,wang2018seismic,konoth2018minesweeper} are proposed for the detection of in-browser cryptojacking malware. There are only a few studies~\cite{darabian2020detecting,caprolu2019crypto} proposed for host-based cryptojacking malware. In addition, Conti et al.~\cite{conti2019detecting}, propose a hardware-level detection mechanism, which can be used to detect both host-based and in-browser cryptojacking malware.

\subsubsection{Analysis Features} \label{analysisfeatures} As can be seen from Table~\ref{tab:detections}, in the cryptojacking domain, the majority of the proposed detection methods are using dynamic analysis. The main reason for this is that mining scripts use a set of known instructions, and they follow and repeat predefined mining steps. For example, miners use cryptographic hash libraries and increment the value of a static variable (i.e., nonce) repeatedly or connect to some known service providers to continue to upload some output results and receive new tasks. These typical behaviors of the cryptojacking malware create a pattern and make them detectable by dynamic analysis. 
In the literature, only a few studies use static features such as opcodes~\cite{darabian2020detecting} and WebAssembly (Wasm) instructions~\cite{konoth2018minesweeper}. WebAssembly~\cite{WASM} is a low-level instruction format that allows programs to run closer to the machine-level language and provide higher performance via stack-based virtual machines~\cite{wasm_blog}. This low-level instruction model lets the WebAssembly run the codes more efficiently, and this feature provides more profit because the cryptojacking script eliminates most of the delay caused by the code execution process. All major browsers in the market currently support WebAssembly. 

Opcodes are machine language instructions that specify the operations to be performed and are used by system calls. The proposed detection system in~\cite{darabian2020detecting} uses opcodes for static analysis, where opcodes are extracted using IDA Pro. In the cryptojacking example, opcodes focus on requests between mining scripts and the operating system's kernel. With this method, they achieve 95\% accuracy with the Random Forest classifier.

On the other hand, many detection mechanisms have been proposed~\cite{rodriguez2018rapid,ning2019capjack,conti2019detecting,kharraz2019outguard,keltonbrowser,neto2020minecap,hong2018you,musch2018web,tahir2019browsers} using dynamics features. The most commonly used dynamic features in these studies are as follows:

\begin{itemize}[leftmargin=*]
    \item \textit{CPU Events~\cite{rodriguez2018rapid,ning2019capjack,kharraz2019outguard,keltonbrowser,musch2018web, yulianto2019mitigation,bian2020minethrottle,petrov2020coinpolice,lachtar2020cross,tanana2020behavior,mani2020decrypto}:} CPU events are the most commonly used features among the dynamic analysis-based detection mechanisms because in-browser cryptojacking scripts have to fetch the CPU instructions to perform the mining, independent of the used hardware. If an in-browser operation uses cryptographic libraries too frequently, which is abnormal for regular websites, it can be directly detected by CPU instructions. Even though CPU is the most crucial feature of cryptocurrency mining, using only CPU events as features may cause high false-positive rates (FPR) because flash gaming or online rendering websites also use the CPU of the system heavily for their operations. To keep FPR as low as possible, most detection methods use more than one features simultaneously~\cite{rodriguez2018rapid,ning2019capjack,keltonbrowser,konoth2018minesweeper,bian2020minethrottle,petrov2020coinpolice,tanana2020behavior,mani2020decrypto}.
    
    \item \textit{Memory activities~\cite{rodriguez2018rapid,conti2019detecting,ning2019capjack,keltonbrowser}:}
   Memory activity is another commonly used feature among the dynamic detection methods listed in Table~\ref{tab:detections}.

\begin{table*}[h]
\centering
\caption{The list of publicly available blacklists.} \label{tab:blacklist}
\begin{tabular}{|c|c|c|} \hline
Ref    & Link   \\ \hline

Nocoin~\cite{NoCoin} & \url{https://github.com/keraf/NoCoin}  \\ \hline

CoinBlocker~\cite{CoinBlockerLists} & \url{https://zerodot1.gitlab.io/CoinBlockerListsWeb/index.html}  \\ \hline

Minerblock~\cite{MinerBlock} &\url{https://github.com/xd4rker/MinerBlock/blob/master/assets/filters.txt}  \\ \hline

Coinhive Blocker~\cite{CoinHiveBlockerGithubPage} &\url{https://raw.githubusercontent.com/Marfjeh/coinhive-block/master/domains}  \\ \hline

Andreas CH Blocker~\cite{andreas} &\url{https://raw.githubusercontent.com/andreas0607/CoinHive-blocker/master/blacklist.json}  \\ \hline

\end{tabular}
\end{table*}

\begin{table*}
\centering
\caption{The list of open-source cryptojacking malware detection implementations.} \label{tab:opensources}
\renewcommand{\arraystretch}{1.0}
\begin{tabular}{|l|c|c|c|}\hline
Ref & Implementation Link & Description & Last Update \\ \hline

CMTracker ~\cite{hong2018you} &  \url{https://github.com/deluser8/cmtracker} &
\makecell{code} & Sep 21, 2018 \\ \hline

Minesweeper~\cite{konoth2018minesweeper} &\url{https://github.com/vusec/minesweeper} & \makecell{data and code} & Mar 17, 2020  \\ \hline

OUTGUARD~\cite{kharraz2019outguard} & \url{https://github.com/teamnsrg/outguard} &
\makecell{data and code} & Sep 6, 2019 \\ \hline
  
SEISMIC~\cite{wang2018seismic} & \url{https://github.com/wenhao1006/SEISMIC} &\makecell{code} & Sep 10, 2019\\ \hline

Retro Blacklist~\cite{holz2020retrospective} & 
\url{https://github.com/retrocryptomining/}  & data and code & Jul 16, 2020
 \\ \hline
\end{tabular}
\end{table*}

\item \textit{Network package~\cite{rodriguez2018rapid,i2019detecting,ning2019capjack,keltonbrowser,neto2020minecap,caprolu2019crypto}:} Network packages are also a handy and useful method to detect cryptojacking activity because of the massive network traffic generated while uploading the calculated hash values to the service provider. The studies~\cite{rodriguez2018rapid,i2019detecting,ning2019capjack,keltonbrowser} utilized network traffic rate as an additional feature along with other features such as memory and CPU-related features. On the other hand, the studies in~\cite{neto2020minecap,caprolu2019crypto} used only network packages for cryptojacking malware detection. Particularly, Neto et al.~\cite{neto2020minecap} use the network flow as a feature, while Caprolu et al.~\cite{caprolu2019crypto} use interarrival times and packet sizes as features in their detection algorithm.

\item \textit{JavaScript (JS) compilation and execution time~\cite{kharraz2019outguard,petrov2020coinpolice}:} In~\cite{kharraz2019outguard,petrov2020coinpolice}, it has been shown that JS engine execution time and JS compilation time is significantly affected by cryptojacking malware. However, online games and other online rendering platforms can also cause the same behavior causing false positives in the detection mechanism. Therefore, the study in~\cite{kharraz2019outguard} also uses CPU usage, garbage collector, and iframe resource loads as secondary features to obtain more accurate results and decrease false positives. The garbage collector is a feature of the JS programming language to optimize memory usage, and it deletes unnecessary data from memory and prevents memory overloading. \ege{The memory} and CPU continuously interact with each other during the mining operation, and the CPU sends calculated data to the memory. The garbage collector deletes all calculated hash values one by one after being sent to the service provider; therefore, the mining process causes irregular usage of the garbage collector. Due to this behavior, the garbage collector can be used as a feature for the dynamic detection mechanism. Iframes are the HTML tags used for embedding another program/function to an HTML source code. Mining scripts are inserted into those tags and work under HTML codes. Similar to previous features, cryptojacking scripts cause irregular usage in iframe resource loads. This feature cannot be used as a primary feature because too many modern web applications use iframe resources irregularly, and it may cause a high false-positive rate.

\item \textit{Hardware Performance Counter (HPC)~\cite{tahir2019browsers,petrov2020coinpolice,mani2020decrypto}:} \ege{HPC values~\cite{das2019sok} are used on modern computers' CPUs and keep
the record of internal CPU events (e.g., Cycles, Cache misses).} The values of the registers with CPU clock cycles and executed instructions provide unique information about the behaviors of a running application. Several studies check the hardware activities and the related applications with HPC values to detect the cryptocurrency mining operations on the system.

\item \textit{System calls~\cite{darabian2020detecting}:} System calls are the API structures that enable the connection between applications and the running system's kernel. System calls run with level 0 privileges to invoke calls and request services from the OS's kernel. The proposed detection system in~\cite{darabian2020detecting} uses the system calls for dynamic analysis, and system calls are recorded using the Cuckoo Sandbox. Then, the system calls are used to train deep learning models, and they achieve 99\% accuracy. 

\end{itemize}

\subsubsection{Classifier and Performance} The collected features are mostly used to train different machine learning classifiers such as Support Vector Machine (SVM)~\cite{rodriguez2018rapid,conti2019detecting,kharraz2019outguard}, Random Forest~\cite{conti2019detecting,tahir2019browsers}, Neural Network~\cite{keltonbrowser,darabian2020detecting}, Decision Tree~\cite{i2019detecting}. Moreover, Neto et al.~\cite{neto2020minecap} proposed the use of incremental learning, which takes the classification probabilities of an ensemble of classifiers as a feature for an incremental learning process. Moreover, Hong et al.~\cite{hong2018you}, proposed a threshold-based detection, and the studies in~\cite{wang2018seismic,konoth2018minesweeper} used a static matching method to detect certain functions in the script. Musch et al.~\cite{musch2019thieves} only report the number of detected websites in the Top 1M Alexa websites. As can be seen from Table~\ref{tab:detections}, all classifiers achieve a near-perfect ($\sim$100\%) detection results.

\subsubsection{Open Source Implementations} Finally, some of the studies \ege{~\cite{hong2018you,konoth2018minesweeper,kharraz2019outguard,wang2018seismic,RetroCryptominingGithubpage}} published their code to help the research community.\ege{Table 3} presents the list of open-source cryptojacking malware implementations.
\vspace{-7pt}
\subsection{Cryptojacking Prevention Studies}\label{sec:prevention}

A majority of the detection mechanisms do not focus on preventing or \ege{interrupting} of cryptojacking malware; however, there are still several studies\cite{yulianto2019mitigation,bian2020minethrottle,razali2019cmblock} focusing on both the detection and prevention of cryptojacking malware. Using dynamic features to detect ongoing cryptojacking is like other dynamic analysis studies, but their prevention methods vary. While Yulianto et al. \cite{yulianto2019mitigation} only raises a notification, Bian et al. \cite{bian2020minethrottle} sleep the mining process, and  Razali et al. \cite{razali2019cmblock} directly \ege{kill} the related process. 

\begin{figure}[h]
    \centering
    \includegraphics[width=.75\columnwidth]{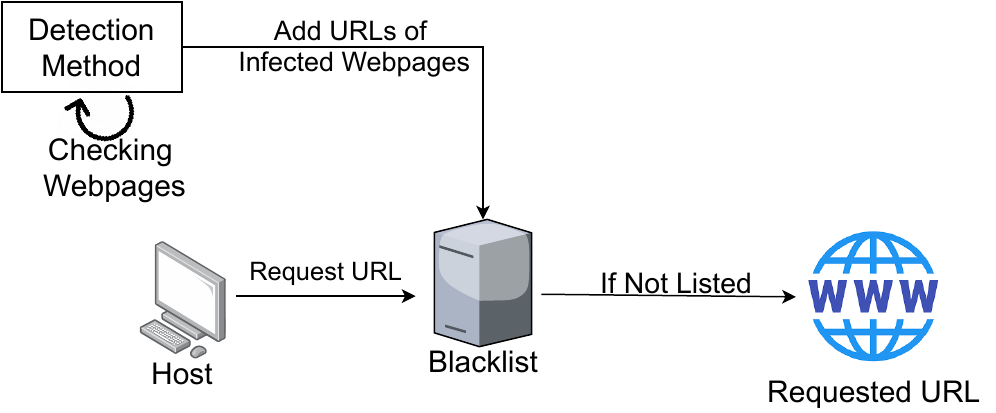}
    \caption{Blacklisting method.}
    \label{fig:blacklisting}
\end{figure}

\vspace{-10pt}

\begin{table*}
\caption{Cryptojacking malware analysis studies in the literature.}
\label{tab:analysis-studies}
\resizebox{\textwidth}{!}{%
\begin{tabular}{|l|l|l|l|}
\hline
\textbf{Ref} & \textbf{Cryptojacking Dataset}     & \textbf{Sample Type}                           & \textbf{Focus of the Study}                                                    \\ \hline
\cite{huang2014botcoin}           & 2000 executable       &   binary         & the practice of using compromised PCs to mine Bitcoin                          \\ \hline
\cite{eskandari2018first}           & 33282 websites     &     script                      & prevalence analysis                                                            \\ \hline
\cite{sigler2018crypto}           & -                              &    -           & how cybercriminals are exploiting cryptomining                                 \\ \hline
\cite{bijmans2019inadvertently}           & 5190 websites         & script         & campaign and domain analysis                                                      \\ \hline
\cite{meland2019experimental}            & XMR-stak, cpuminer-multi  &    binary                & attack impact on consumer devices and user annoyance                           \\ \hline
\cite{saad2019dine}           & 5700   websites                &   script            & static, dynamics and economic analysis                                         \\ \hline
\cite{zimba2019cryptojacking}           & CoinHive   cryptominer    &     script              & sample characteristics  and network traffic analysis                           \\ \hline
\cite{pastrana2019first} & 1.2   million miners & binary &  currencies,   actors , campaign and earning analysis, underground markets \\ \hline
\cite{papadopoulos2019truth}           & 107511   websites       &   script                   & profitability and the imposed overheads                                        \\ \hline
\cite{bijmans2019just}           & 3.2 TB historical scan results      &   script         & investigation of a new type of attack that exploits Internet infrastructure for cryptomining                        \\ \hline
\cite{carlin2019you}            & -            &       -                          & business model, threat sources, implications, mitigations, legality and ethics \\ \hline
\cite{aziz2020coinhive}           & 53 websites     &       script                       & sample characteristics                                                         \\ \hline
\cite{varlioglu2020cryptojacking}            & 2770   websites   &   script                         & activeness analysis                                                            \\ \hline
\cite{zimba2020crypto}           & XMRig miner       &    binary                        & sample characteristics                                                         \\ \hline
\cite{holz2020retrospective}            &156 domains, 25892 proxies & script &  impact on the web users                                                        \\ \hline
\end{tabular}%
}
\end{table*}

For cryptojacking prevention, there are also several tools in the market. Against host-based cryptojacking malware, proprietary antivirus programs~\cite{AvastAntiMAlware, Norton}\footnote{As these programs are closed-source, their methods are not publicly available.} are commonly preferred. Against in-browser cryptojacking malware, open-source browser extensions such as NoCoin~\cite{NoCoin} and MinerBlock~\cite{MinerBlock} are widely used. These open-source browser extensions are based on blacklisting, where the lists are updated as new malicious domains are discovered. Table~\ref{tab:blacklist} shows the list of publicly available blacklists that we identified during our research. Browser extensions warn the user when the user wants to access a  website on the blacklist. Figure~\ref{fig:blacklisting} shows the blacklisting process, which is repeated as a continuous loop.

Pure blacklisting-based prevention is not an efficient way for stopping cryptojacking malware because attackers can easily change their domain by domain fluxing or other methods to downshift the effects of blacklists. There are also some new methods~\cite{ramanathanblag} proposed by researchers for better and more optimized blacklisting, but even dynamic blacklisting methods are not fully effective nor protective~\cite{yadav2012detecting} against domain fluxing methods.

\subsection{Cryptojacking Analysis Studies}\label{sec:anaylsis}
In addition to the cryptojacking malware detection and prevention studies, some researchers also performed empirical measurement studies to understand the cryptojacking threat landscape better. \abbas{Table \ref{tab:analysis-studies} shows cryptojacking malware analysis studies in the literature.} In these studies, \abbas{cryptominers are either in the format of} binary \cite{meland2019experimental,zimba2020crypto,pastrana2019first,huang2014botcoin} \abbas{or} script \cite{bijmans2019inadvertently,holz2020retrospective,saad2019dine,varlioglu2020cryptojacking,eskandari2018first,zimba2019cryptojacking,aziz2020coinhive,papadopoulos2019truth}
except \cite{carlin2019you,sigler2018crypto} where the findings in these studies are based on the other studies and publicly available documents.

Researchers analyzed several different perspectives of cryptojacking. In the first study \cite{huang2014botcoin}, the authors analyze the binary samples identified as engaged in mining operations to characterize their scope, operations, and revenue. This is the first and only study analyzing Bitcoin \ege{miners}, where the samples used in other studies are mining Monero. The increase in the cryptojacking malware attack instances in 2017 also drew researchers' attention. \cite{eskandari2018first} is the first study analyzing the Monero cryptojacking samples, where the authors used over 30000 websites utilizing \textit{coinhive.min.js} library for the prevalence analysis of cryptojacking samples. Many follow-up studies are published. For example, the studies \cite{zimba2019cryptojacking,aziz2020coinhive,zimba2020crypto} also performed an analysis of the cryptojacking samples to identify characteristics of the samples. In addition, the studies in \cite{meland2019experimental,holz2020retrospective} performed the impact analysis. Particularly, \cite{meland2019experimental}   analyzed the attack impact on consumer devices and user annoyance, and \cite{holz2020retrospective}  analyzed the impact of cryptojacking malware on web users, while \cite{papadopoulos2019truth} analyzed the overhead of cryptojacking samples. In an interesting study, the authors in \cite{bijmans2019just} investigated a new type of attack exploiting the Internet infrastructure for cryptomining, which indeed has an impact on 1.4M infected routers.

Moreover, there are also studies performing the economic analysis of cryptojacking samples such as \cite{saad2019dine,pastrana2019first,papadopoulos2019truth}. Other than that, the authors in \cite{bijmans2019inadvertently,pastrana2019first} performed a campaign analysis of the cryptojacking samples and \cite{varlioglu2020cryptojacking} analyzed the activeness of cryptojacking threat after the discontinuation of Coinhive. Finally, while \cite{sigler2018crypto} gives an overview of how cybercriminals are exploiting cryptomining,  \cite{carlin2019you} presents a review of the business model, threat sources, implications, mitigations, legality, and ethics of cryptojacking malware.

\section{Lessons Learned and Research Directions} \label{sec:research_directions}
\vspace{-6pt}
This section covers lessons we learned during this research and potential research directions that can further be explored by other researchers:

\vspace{2pt}
\noindent \textbf{A recent trend in cryptojacking attacks.}
\noindent \textbf{\deleted{The trend shift from in-browser to host-based cryptojacking attacks.}} \deleted{The discontinuation of Coinhive's service in March of 2019 led to a big drop in the number of in-browser cryptojacking instances observed in the wild} 
\deleted{\deleted{\cite{checkpointreport}}}\deleted{ as it is the leading distributor of cryptojacking scripts and the connection between the mining pools and users are lost.} \deleted{We also observe and verify this finding using both VT and PublicWWW datasets. In the VT dataset, we observe that 84\% of the samples are uploaded right after Coinhive started its service in 2017. However, in 2019 and 2020, the number of samples uploaded to VT is around 14\% in total. Moreover, we also observe a similar result in PublicWWW, in which we can directly see the service provider of the sample. 50\% of the samples with a known service provider (2667/5328) and 43\% (2667/6269) of all samples are still using either Coinhive or Authedmine, which is an obfuscated script service of Coinhive. Both of these observation shows that the alternative service providers introduced after Coinhive did not gain popularity as much as Coinhive, even though we have seen some samples containing the scripts from them. Finally, the actions taken by the platforms such as Google} \deleted{ \deleted{\cite{google,google_apps}}, \deleted{Apple \cite{apple}}, \deleted{Opera \cite{opera}} also made the cryptocurrency mining in the browser and mobile devices less popular, which led the attackers to look for new targets.}

\deleted{On the other hand, s}\added{S}ome security reports published in 2020 \cite{IBMreport2020,cloud-cpr-report} noted 
a\deleted{nother} trend \deleted{shift} in the cryptojacking attacks, in which the attackers now target the devices with more processing power rather than the personal computers as in the in-browser cryptojacking attacks. With this, the attackers' goal is to obtain more profit in a lesser time. Some examples of these targeted devices are enterprise cloud infrastructure \cite{AWS-Hack,aws-credentials}, servers~\cite{Servers850000}, a large number of inadequately protected IoT devices \cite{JackingIoTReport} or Docker engines \cite{watchdog}. In these attacks, the attackers did not only use the Coinhive's script but also modified \added{a} non-malicious and open-source Monero miner called XMRig to perform the cryptomining in the background~\cite{XMRRigDistribute}. Unlike in-browser cryptominers, the client does not come to the attacker; therefore, the attacker needs to deliver the malicious mining script to the victims. For this purpose, the attackers used the vulnerabilities as in the case of Mikrotik routers~\cite{MikrotikNews} or a recent CVE to deliver Monero cryptominer~\cite{cve_monero}, poorly configured IoT devices~\cite{JackingIoTReport}, or poor security~\cite{watchdog}. It is also seen that insiders may want to take advantage of the servers~\cite{russian-nuclear}. 

However, despite the decrease in the number of in-browser samples from active service providers and the potential trend shift in the attackers' behavior to host-based cryptojacking malware and techniques used to deliver the malware, host-based cryptojacking malware literature is not as rich as in-browser cryptojacking malware literature. As can be seen from Table~\ref{tab:detections} and~\ref{tab:analysis-studies}, there are only several studies on the detection~\cite{tanana2020behavior,darabian2020detecting,caprolu2019crypto,lachtar2020cross,ahmad2019new,mani2020decrypto,conti2019detecting} and the analysis \cite{huang2014botcoin,meland2019experimental,pastrana2019first,zimba2020crypto}. 
Therefore, there is a need for more effort from the security researchers to find better solutions to detect and mitigate this continually evolving threat.

\noindent \textbf{\abbas{Monero as a target cryptocurrency.}} \abbas{In recent attacks, Monero has become a de-facto cryptocurrency for the cryptojacking attacks. Another pattern we spotted is that in almost all of the attack instances in the previous section \cite{cloud-docker,watchdog,cloud-Azure,aws-credentials,hidegard,DrivebyMiner,DOD}, the attackers use Monero as a target cryptocurrency instead of Bitcoin or other cryptocurrencies. Even we are not sure about their motive, Monero is the most popular privacy coin hiding the track of the transactions. For example, if the attackers would use Bitcoin, even though the attack has been detected after a long time, it would be possible to track down the Bitcoin transactions.}

\vspace{5pt}
\noindent \textbf{The evaluation of the proposed solutions.} We identified threes issues regarding to the evaluation of the proposed solutions in the literature:

\begin{itemize} [leftmargin=0.25in]
    \item \textit{Dataset dates.} 
    The effectiveness of an in-browser cryptojacking malware detection mechanism is directly related to the number of websites detected. However, the infected websites modify or move their script to other domains frequently to avoid being blacklisted. Moreover, many websites discontinued mining after the Coinhive shutdown~\cite{iscryptojackingdeadaftercoinhive}. Therefore, the accuracy of a detection method may significantly vary depending on when the dataset was collected. \ege{Only five of the proposed detection mechanism \cite{kharraz2019outguard, hong2018you,konoth2018minesweeper,wang2018seismic,holz2020retrospective} in Table~\ref{tab:detections} reports the dataset date. Therefore, \ege{we do not know most of the studies'} dataset collection date; which makes a fair comparison difficult.}

    \item \textit{Online vs. offline detection.} The detection mechanisms proposed in the literature usually focus on accuracy as an evaluation metric, and they mostly claim a near-perfect accuracy in detecting cryptojacking malware. However, most of the time, they do not report how their method was implemented, that is, whether it was offline or online. In offline detection, the sample is detected randomly and added to the database (e.g., signature, blacklist). In the online detection, the sample is detected in a real-time manner.  As it has been shown that detection ratio may vary for online and offline detection \cite{online_vs_offline},
    it is critical for the detection studies to report if the proposed method is implemented in an online or offline environment.
  
    \item \textit{Overhead analysis.} \ege{Only the authors of the two\cite{kharraz2019outguard,konoth2018minesweeper} proposed dynamic analysis tools consider the usability of their detection mechanisms on the end-user side.} But, especially for machine learning-based detection methods, using behavioral features may introduce a high overhead on the end-user side. This should be taken into consideration by researchers in future studies. 
\end{itemize}

\vspace{5pt}
\noindent \textbf{The legitimate use of \abbas{in-browser} cryptocurrency mining.} An issue we identified during our research is that the in-browser cryptocurrency mining was initially started to provide an alternative revenue to the legitimate website owners such as new publishers~\cite{salon} or non-profit organizations like UNICEF~\cite{UnicefMining}. Later, some service providers such as Coinimp~\cite{Coinimp}, WebMinePool~\cite{webminepool} even provided methods for explicit user consent in their implementations. However, with the keyword-based automatic detection and prevention methods such as browser extensions~\cite{NoCoin,MinerBlock} or even browsers themselves~\cite{opera,google} blocking the websites containing cryptomining script, this use of web-based cryptomining scripts is not possible anymore. A practical solution to this issue would be asking for the user's explicit consent instead of directly blocking a website trying to upload a mining script. Moreover, there is a need for more effort by researchers to work on the usage of legitimate cryptomining with user consent and knowledge as a funding model.

\vspace{5pt}
\noindent \textbf{The use of traditional malware attacks on Bitcoin and blockchain infrastructure.} There are two types of Bitcoin- and blockchain-related malware seen in the wild: those that use the Bitcoin and blockchain infrastructure to exploit the victim; or those that use the traditional malware attacks such as key stealing, social engineering, or fake application attacks to exploit Bitcoin and blockchain users. Cryptojacking attacks use the Bitcoin and blockchain infrastructure to exploit the victim's computational power; however, Bitcoin and blockchain users are also exposed to many traditional malware attacks. These attacks specifically aim to obtain Bitcoin and blockchain users' private keys through social engineering methods~\cite{PhishingAttack,PhishingAttackIsrael,PhishingAttackXRP}, fake wallets~\cite{Fake3wallet,Googleplayfakewallet}, and key-stealing trojan malware~\cite{FraudMalware,GoogleChromeextension,Cryptocurrencystealinglandscape}. Although these attacks and their countermeasures~\cite{acar2020usable,acar2019maf,celik2019curie} have been studied extensively in the literature~\cite{heartfield2015taxonomy}, their impact in the Bitcoin and blockchain domain has not been investigated yet and can lead to new research directions.

    
\vspace{-10pt}
\section{Conclusion} \label{sec:conclusion}
The rapid rise of cryptocurrencies incentivized the attackers to the lucrative blockchain and the Bitcoin ecosystem. With ready-to-use mining scripts offered easily by service providers (e.g., Coinhive~\cite{Coinhive}, and CryptoLoot~\cite{crypto-loot}) and untraceable cryptocurrencies (e.g., Monero), cryptojacking malware has become an essential tool for hackers. The lack of mitigation techniques in the market led to many cryptojacking malware detection studies proposed in the literature. In this paper, we first explained the cryptojacking malware types and how they work in a systematic fashion. Then, we presented the techniques used by cryptojacking malware based on the previous research papers, cryptojacking samples, and major attack instances. In particular, we presented sources of cryptojacking malware, infection methods, victim platform types, target cryptocurrencies, evasion, and obfuscation techniques used by cryptojacking malware. Moreover, we gave a detailed review of the existing detection and prevention studies as well as the cryptojacking analysis studies in the literature. Finally, we presented lessons learned, and we noted several promising new research directions. In doing so, this \ege{SoK} study will facilitate not only a deep understanding of the emerging cryptojacking malware and the pertinent detection and prevention \ege{mechanisms but} also a substantial additional research work needed to provide adequate mitigations in the community.

\vspace{-10pt}
\section*{Acknowledgment} \label{sec:acks}
\vspace{-5pt}
We would like to thank VirusTotal for sharing the samples. We also would like to thank the anonymous reviewers, and our shepherd Dr. Christian Rossow for their feedback and time. This work was partially supported by the U.S. National Science Foundation (NSF) (Awards: NSF-CAREER CNS-1453647, NSF-1663051, NSF-CNS-1718116, NSF-CNS-1703454), and ONR under the "In Situ Malware" project, and CyberFlorida Capacity Building Program. The views expressed are those of the authors only.

\bibliographystyle{IEEEtran}
\bibliography{refs.bib}

\appendices
\vspace{-10pt}
\section{\abbas{Papers Distribution}} \label{sec:paper_distribution}
\vspace{-5pt}

\begin{figure}[h]
    \centering
    \includegraphics[width=.65\columnwidth,trim=0cm 0.5cm 0cm 0cm]{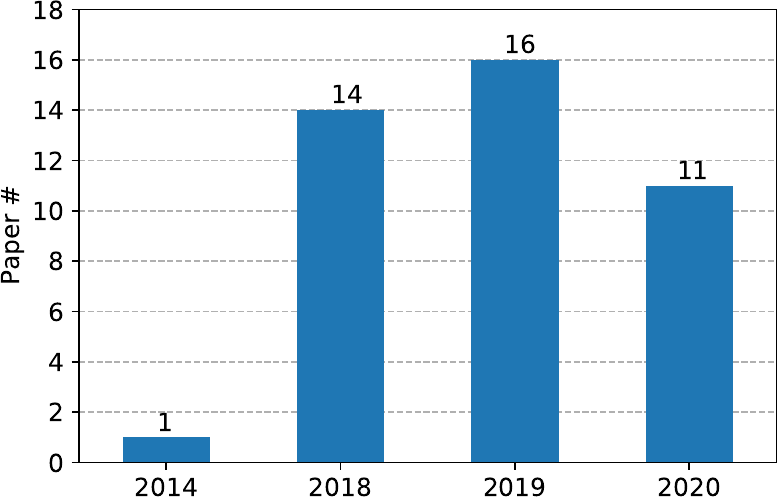}
     \caption{Yearly distribution of 42 cryptojacking-related academic research papers we used in this study.}
    \label{fig:paperdistribution}
\end{figure}

\abbas{Figure \ref{fig:paperdistribution} shows the distribution of cryptojacking-related research papers per year.}

\vspace{-10pt}

\section{\abbas{Samples Distribution}}\label{sec:samples_distribution}
\vspace{-5pt}
In this section, our goal is to give more details about the VT and PublicWWW datasets, perform quantitative and longitudinal analysis on our two datasets to confirm some of our findings in the paper, \added{present the limitations of the datasets,} and give more insights on the dataset. More details about both VT and PublicWWW datasets can be found \deleted{in} \added{on} the following website: \url{https://github.com/sokcryptojacking/SoK}
\vspace{-5pt}
\subsection{\abbas{VT Dataset}} \label{subsec:vt_dataset}
\vspace{-5pt}
We used VT Academic API to access the VT dataset consisting of 437279 unique samples (both cryptojacking and non-cryptojacking) and their VT scan reports in the format of JSON. To detect the cryptojacking samples among all samples, we used AV labels in the scan reports of these samples and looked for the keyword "miner", i.e., if any of Antivirus (AV) label in the report include the keyword "miner", we included in our samples. We picked the keyword "miner" as we considered it to be the most generic keyword to find all of the cryptojacking samples, and it is also used in recent work as a generic class label for the VT samples~\cite{acsac2020_avlabel}. Our scan resulted in the 20200 cryptojacking malware samples. \added{We want to note that this method for selecting cryptojacking samples will not detect the samples that AVs can not label.}

\subsubsection{\abbas{More Insights on VT Dataset}}
\abbas{In addition to the AV labels that we used to detect the miners, VT scan reports also include other information regarding the samples such as \textit{first seen} date, file type, submission names, the total number of detection by AVs of the samples. We performed more analysis using this information and explain our results in the rest of the section.}

\begin{figure}[h]
    \centering
    \includegraphics[width=.65\columnwidth,trim=0cm 0.5cm 0cm 0.5cm]{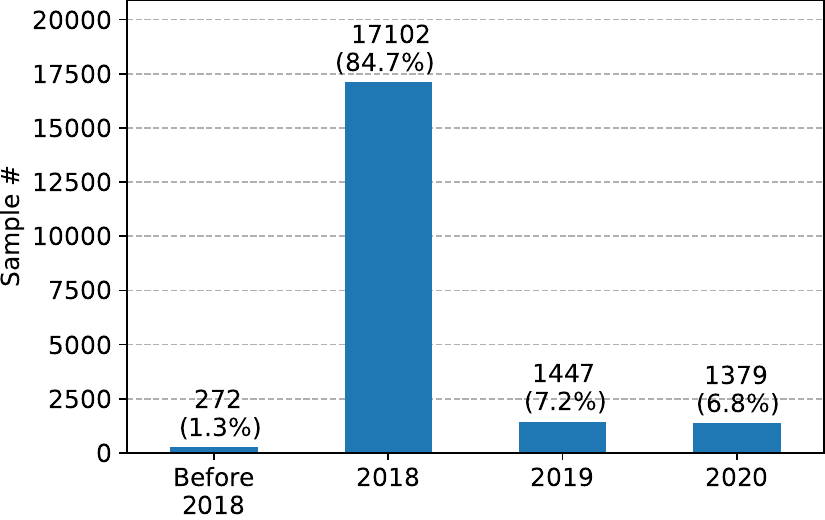}
     \caption{Yearly distribution of VT dataset.}
    \label{fig:VT_dataset_per_year}
\end{figure}

\noindent \textbf{\deleted{Yearly}\added{Time} Distribution:} Figure \ref{fig:VT_dataset_per_year} shows the yearly distribution of the samples' first seen date in the \ege{VT academic} dataset \cite{vt_first_seen}. \deleted{As can be seen from the figure, only 1\% of the cryptominer samples were submitted to VT before 2018. This indeed matches and verifies the expected finding of the surge in the number of cryptominers after Coinhive started its service in 2017. Moreover, the sharp decrease after 2018 is also notable. This may indicate that not many new samples are found in the wild in 2019 and 2020; however, it does not show that the number of active samples in 2019 and 2020 is less than in 2017 as the samples uploaded in 2017 may be still actively operating. These samples' activeness requires more analysis on the script used in those samples and real-time capture of the samples in the wild.}

\vspace{-5pt}
When we continued more \deleted{granular time} analysis on the samples, \added{we found the VT dataset has the following two limitations:} 
\begin{itemize}
    \item There are more cryptojacking samples that can be accessed through the VT interface with more privileges. Therefore, this dataset may have bias in terms of the representation of the real-life cryptojacking samples\lastdeleted{as it only covers the subset of the samples uploaded to the VT}. \lastadded{For example, as can be seen from Figure \ref{fig:VT_dataset_per_year}, $85\%$ of all cryptojacking samples in the VT dataset are from 2018; therefore, the samples from 2018 are over-represented in the VT dataset.}
    \item \deleted{we found that s}\added{S}amples \added{in the VT dataset} are uploaded to VT as a batch every six months. Therefore, we concluded that a smaller time frame analysis than the yearly distribution might not be reliable as representing the time distribution of real-life samples seen in the wild.
\end{itemize}

\begin{figure}[h]
    \centering
    \includegraphics[width=.65\columnwidth,trim=0cm 0.5cm 0cm 0cm]{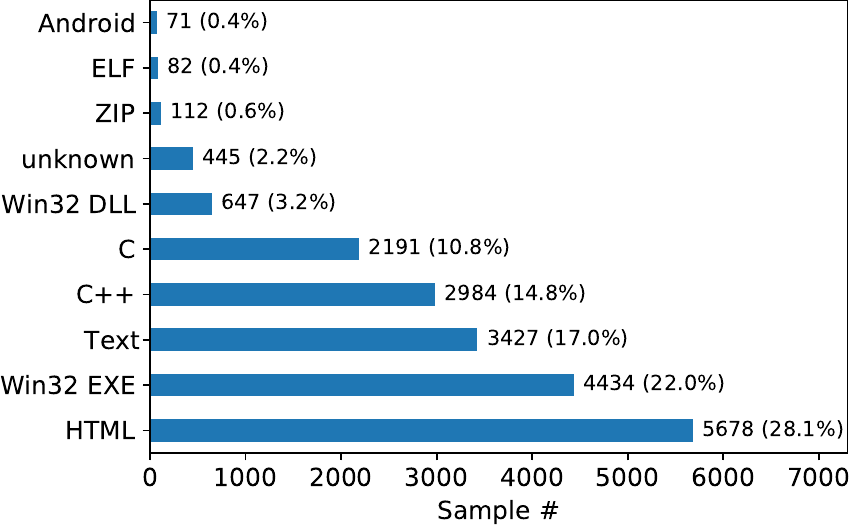}
     \caption{Top-10 file types of the samples in VT dataset.}
    \label{fig:VT_dataset_file_type}
\end{figure}

\noindent \abbas{\textbf{File Type Distribution:}} \abbas{In order to detect the file type, we used the type given by the VT scan reports \cite{vt_first_seen}. Figure \ref{fig:VT_dataset_file_type} shows the top 10 file types of the samples in the VT dataset. According to the figure, HTML is the most common file type in the VT dataset, while the Win32 EXE is the second most common file type.}

\vspace{-5pt}

\abbas{File type distribution of VT dataset samples is important as they can be used to decide if the sample is an in-browser or host-based type of cryptojacking. Even though some of the file types clearly indicate the type of the cryptojackings, some may require a more in-depth analysis of the sample. In-browser samples only contain the mining script and in the form of text format to embed in the website source code, while the host-based cryptojacking malware samples are in the executable or other formats that can be run on the host machine. For example, we found that all HTML files are in-browser samples while Win32 EXE and Win32 DLL samples are host-based cryptojacking samples. However, for the file types such as Text, C, C++, ZIP, one needs to check the sample itself and other useful information like submission names to decide whether the sample is in-browser or host-based. }

\begin{figure}[h]
    \centering
    \includegraphics[width=.65\columnwidth,trim=0cm 0.5cm 0cm 0cm]{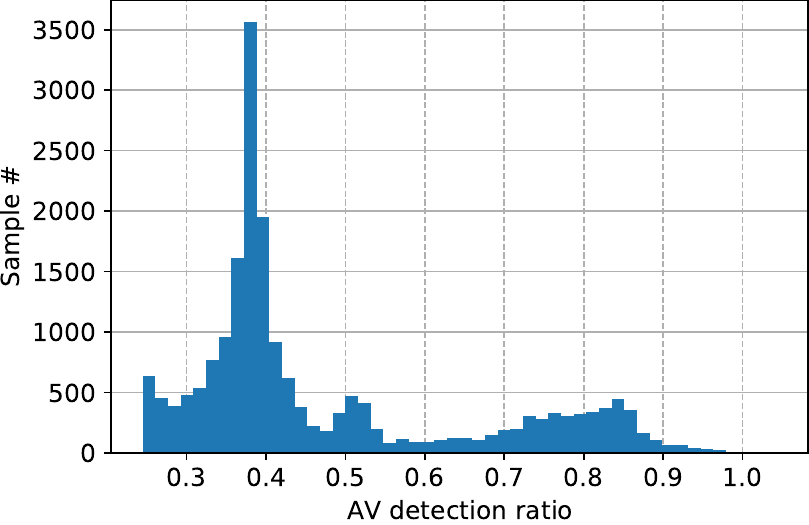}
     \caption{Histogram distribution of detection ratio of the samples in VT dataset.}
    \label{fig:VT_dataset_detection_ratio}
\end{figure}

\noindent \abbas{\textbf{Detection Ratio:}} \abbas{VT scan reports includes the detection results of around 60 vendors for each sample \cite{vt_first_seen}. In this part, our goal is to able to see the detection ratio of AVs for the cryptojackings samples in our VT dataset. For this, we plot the histogram of the detection ratio, and the results are given in Figure \ref{fig:VT_dataset_detection_ratio}.}

\vspace{-5pt}
\abbas{The results show that the average detection ratio of cryptojacking samples is approximately 40\%. We note that every cryptojacking sample in our dataset is detected by at least one AV vendor. This is because of the filtering method we used to find the cryptojacking samples among all samples, i.e., searching for the keyword "miner" among all AV labels. Therefore, if any AV vendors have not detected a sample, it would not be in our dataset in the first place. This is a limitation of the VT dataset. In order to overcome this limitation and create a recent and more comprehensive dataset, we created another dataset, which we will explain in the next section.}

\subsection{\abbas{PublicWWW Dataset}}
\vspace{-5pt}
\abbas{The VT dataset does not include the samples that can bypass the AV detection methods and samples that have never submitted to VT. In order to create a more comprehensive and recent dataset of cryptojacking malware, we used the HTML source keyword search engine PublicWWW~\cite{publicwww}. We created the PublicWWW dataset using the following steps: 
\begin{enumerate}[leftmargin=0.15in]
    \item We obtained the keyword lists from the blacklists~\cite{NoCoin,MinerBlock}, previous studies~\cite{bijmans2019inadvertently,konoth2018minesweeper} and manual analysis of the samples from the VT dataset. Particularly, we used a merged blacklist from NoCoin~\cite{NoCoin} and MinerBlock~\cite{MinerBlock}; 76 keywords from~\cite{bijmans2019inadvertently} and 38 keywords from~\cite{konoth2018minesweeper}; 25 keywords from the VirusTotal samples.
    \item We downloaded the list of URLs for each keyword from PublicWWW.
    \item We merged the lists and removed the duplicates to obtain a unique list of URLs.
    \item We used a web crawler to download the HTML source code of each URL.
    \item We verified the samples by checking the keywords in their source code and removed the samples that do not satisfy this condition.
\end{enumerate}
 This process resulted in 6269 unique URLs, their HTML source codes, and their final keyword list with 154 unique keywords used in these samples. From the previous two studies~\cite{bijmans2019inadvertently,konoth2018minesweeper} and our findings of publicly known service providers, we identified 14 service providers in total. We manually analyzed their documentation and found that 5328 samples are using the scripts from those 14 service providers. Then, we identified 24 unique keywords to uniquely capture the samples. We released the service provider and keywords lists in our dataset link.}


\vspace{-5pt}
\begin{figure}[h]
    \centering
    \includegraphics[width=.85\columnwidth,trim=0cm 0.5cm 0cm 0cm]{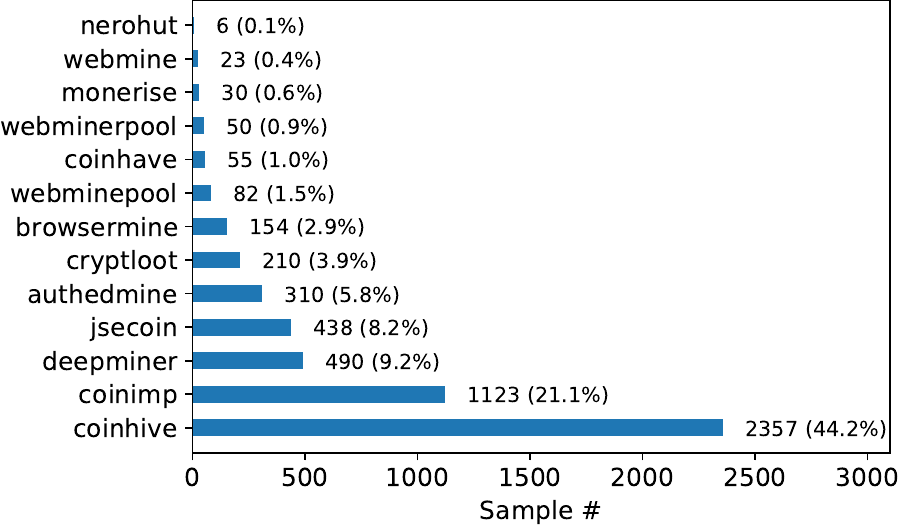}
     \caption{Service provider distribution of the samples in PublicWWW dataset.}
    \label{fig:service_provider_dist}
\end{figure}
\vspace{-5pt}

\abbas{Figure \ref{fig:service_provider_dist} shows the service provider distribution of the samples in the PublicWWW dataset. As shown in the figure, even though it is inactive, Coinhive is still the most common service provider among all. On the other hand, Coinimp is the second highest service provider and it is still active as of writing this paper. In addition, we found that 144 samples are using scripts from the multiple samples while we are not able \ege{to identify} the associated service provider of 941 samples, which we marked as domain lists with \ege{an unknown} service provider.} \added{We also want to note that the samples we have in the PublicWWW are captured during this paper's experiments, but it does not mean that these domains will contain the cryptocurrency mining script any time in the future. Therefore, one may need to re-verify the existence of cryptocurrency mining scripts for their analysis by checking the source code.}

\vspace{-5pt}
\section{\abbas{Attack Instances Distribution}} \label{sec:attacks_distribution}
\vspace{-5pt}
\begin{figure}[h]
    \centering
    \includegraphics[width=.65\columnwidth,trim=0cm 0.5cm 0cm 0cm]{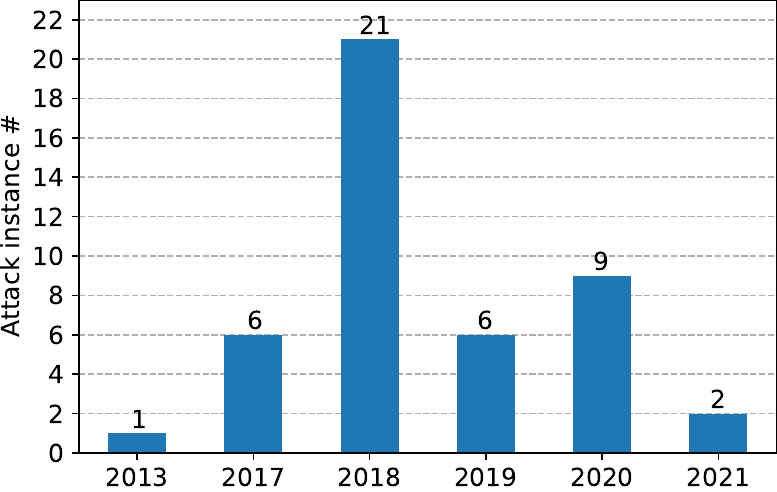}
     \caption{Yearly distribution of 39 cryptojacking attacks instances we used in this paper.}
    \label{fig:attack_distribution}
\end{figure}
\vspace{-5pt}

\abbas{Figure \ref{fig:attack_distribution} shows the distribution of the attack instances we used in this paper per year. This distribution graph does not show any indicative result regarding the cryptojacking malware's popularity over time in our paper. 
Only one attack instance from 2013 may seem like an outlier; however, that example shows one of the first instances of cryptojacking malware idea, which is very similar to its usage after 2018. In that attack, a cryptojacking malware attack is instantiated by attaching the sample inside a video game to mine Bitcoin. Finally, in addition to 45 major attack instances, we also added 14 service providers' webpage and 5 blacklists' link and shared in our dataset link.}

\end{document}